\documentclass[twocolumn, secnumarabic, amssymb, aps, prb, superscriptaddress, 10pt]{revtex4-1}
\usepackage{graphicx}
\usepackage{float}
\usepackage{bm}
\usepackage{braket}

\usepackage{xspace}

\renewcommand{\approx}{\simeq}
\usepackage{amsmath}

\usepackage{bbm}
\newcommand{\equref}[1]{Eq.~(\ref{#1})}

\newcommand{\secref}[1]{Sec.~\ref{#1}}
\newcommand{\figref}[1]{Fig.~\ref{#1}}
\newcommand{\refcite}[1]{Ref.~\onlinecite{#1}}
\newcommand{\refscite}[1]{Refs.~\onlinecite{#1}}

\newcommand{\appref}[1]{Appendix~\ref{#1}}
\renewcommand{\vec}[1]{\boldsymbol{#1}}
\newcommand{\pdagger}{{\phantom{\dagger}}}

\usepackage{color}
\usepackage[normalem]{ulem}

\begin{document}

\title{Electron irradiation effects on superconductivity in PdTe$_2$: \\ an application of a generalized Anderson theorem}

\author{E.~I.~Timmons}
\affiliation{Ames Laboratory, Ames, Iowa 50011, USA}
\affiliation{Department of Physics and Astronomy, Iowa State University, Ames, Iowa 50011, USA }	

\author{S. Teknowijoyo}
\affiliation{Ames Laboratory, Ames, Iowa 50011, USA}
\affiliation{Department of Physics and Astronomy, Iowa State University, Ames, Iowa 50011, USA }	

\author{M. Ko\'nczykowski}
\affiliation{Laboratoire des Solides Irradi\'es, \'Ecole Polytechnique, CNRS, CEA, Universit\'e Paris-Saclay, 91128 - Palaiseau Cedex, France}

\author{O. Cavani}
\affiliation{Laboratoire des Solides Irradi\'es, \'Ecole Polytechnique, CNRS, CEA, Universit\'e Paris-Saclay, 91128 - Palaiseau Cedex, France}

\author{M.~A.~Tanatar}
%\email{tanatar@ameslab.gov}
\affiliation{Ames Laboratory, Ames, Iowa 50011, USA}
\affiliation{Department of Physics and Astronomy, Iowa State University, Ames, Iowa 50011, USA }

\author{Sunil~Ghimire}
\affiliation{Ames Laboratory, Ames, Iowa 50011, USA}
\affiliation{Department of Physics and Astronomy, Iowa State University, Ames, Iowa 50011, USA }

\author{Kyuil Cho}
%\email{kcho@ameslab.gov}
\affiliation{Ames Laboratory, Ames, Iowa 50011, USA}

\author{Yongbin Lee}
\affiliation{Ames Laboratory, Ames, Iowa 50011, USA}

\author{Liqin Ke}
\affiliation{Ames Laboratory, Ames, Iowa 50011, USA}

\author{Na~Hyun~Jo}
%\email{njo@iastate.edu}
\affiliation{Ames Laboratory, Ames, Iowa 50011, USA}
\affiliation{Department of Physics and Astronomy, Iowa State University, Ames, Iowa 50011, USA }

\author{S.~L.~Bud'ko}
%\email{budko@ameslab.gov}
\affiliation{Ames Laboratory, Ames, Iowa 50011, USA}
\affiliation{Department of Physics and Astronomy, Iowa State University, Ames, Iowa 50011, USA }

\author{P.~C.~Canfield}
%\email{canfield@ameslab.gov}
\affiliation{Ames Laboratory, Ames, Iowa 50011, USA}
\affiliation{Department of Physics and Astronomy, Iowa State University, Ames, Iowa 50011, USA }

\author{Peter~P.~Orth}
%\email{porth@iastate.edu}
\affiliation{Ames Laboratory, Ames, Iowa 50011, USA}
\affiliation{Department of Physics and Astronomy, Iowa State University, Ames, Iowa 50011,
USA }

\author{Mathias~S.~Scheurer}
%\email{mscheurer@g.harvard.edu}
\affiliation{Department of Physics, Harvard University, Cambridge, Massachusetts 02138, USA}

\author{R.~Prozorov}
\email{Corresponding author: prozorov@ameslab.gov}
\affiliation{Ames Laboratory, Ames, Iowa 50011, USA}
\affiliation{Department of Physics and Astronomy, Iowa State University, Ames, Iowa 50011, USA }

\date{\today}

\begin{abstract}
Low temperature ($\sim$ 20~K) electron irradiation with 2.5~MeV relativistic electrons was used to study the effect of controlled non-magnetic disorder on the normal and superconducting properties of the type-II Dirac semimetal PdTe$_2$. We report measurements of longitudinal and Hall resistivity, thermal conductivity and London penetration depth using tunnel-diode resonator technique for various irradiation doses. The normal state electrical resistivity follows Matthiessen rule with an increase of the residual resistivity at a rate of $\sim$0.77$ \mu \Omega$cm/$(\textrm{C}/\textrm{cm}^2)$. London penetration depth and thermal conductivity results show that the superconducting state remains fully gapped. The superconducting transition temperature is suppressed at a non-zero rate that is about sixteen times slower than described by the Abrikosov-Gor'kov dependence, applicable to magnetic impurity scattering in isotropic, single-band $s$-wave superconductors. In order to gain information about the gap structure and symmetry of the pairing state, we perform a detailed analysis of these experimental results based on insight from a generalized Anderson theorem for multi-band superconductors. This imposes quantitative constraints on the gap anisotropies for each of the possible pairing candidate states. We conclude that the most likely pairing candidate is an unconventional $A_{1g}^{+-}$ state. While we cannot exclude the conventional $A_{1g}^{++}$ and the triplet $A_{1u}$, we demonstrate that these candidates require additional assumptions about the orbital structure of the disorder potential to be consistent with our experimental results, e.g., a ratio of inter- to intra-band scattering for the singlet state significantly larger than one. Due to the generality of our theoretical framework, we believe that it will also be useful for irradiation studies in other spin-orbit-coupled multi-orbital systems.
\end{abstract}

\maketitle

\section{Introduction}
\label{sec:introduction}
The layered transition metal dichalcogenide (TMD) superconductor PdTe$_2$~\cite{Bell,Canadian} has received renewed interest recently after the discovery of type-II Dirac points in its bulk band structure~\cite{Noh-2017}. As shown by angle-resolved photo-emission (ARPES) and density-functional theory (DFT) calculations~\cite{dHvA,PdTe-ARPES}, the nodal points, which are protected by three-fold rotation symmetry, lie about $0.6$~eV below the Fermi energy $E_F$ and occur along the $\Gamma$-A line in the Brillouin zone. The Fermi surface consists of several electron pockets around the K and K$'$ points and two hole pockets around $\Gamma$, whose energy bands eventually cross at the Dirac point. A proposed mechanism that can explain the occurrence of this crossing~\cite{Bahramy-NatMater} invokes a band inversion of chalcogenide $p$ states in this strongly spin-orbit coupled material and was shown to be relevant for a number of other TMDs as well. 

PdTe$_2$ becomes superconducting below a transition temperature of $T_c = 1.7$~K~\cite{Bell,Canadian}. The superconducting state was consistently found to be fully gapped in a number of experiments performing thermodynamic~\cite{type1SC}, penetration depth~\cite{PRB}, scanning tunnel microscopy (STM)~\cite{STM}, and heat capacity~\cite{HC} measurements. Superconductivity was reported to be of type-I based on magnetization ~\cite{type1SC} and muon spin rotation~\cite{type1SCmuSR} studies. This is consistent with an experimentally observed Ginzburg ratio of $\kappa = \xi/\lambda \approx 0.52 < 1/\sqrt{2}$, when $\xi$ and $\lambda$ are directly extracted from critical field $H_c(0)$~\cite{type1SC} and penetration depth~\cite{PRB} measurements, respectively. 

The presence of strong spin-orbit coupling and band inversions provide a natural motivation for a detailed investigation of the symmetry of the superconducting pairing state in this multi-band system. In a previous work, Teknowijoyo \emph{et al.}~\cite{PRB} have performed a systematic classification of all possible translationally invariant superconducting pairing states in PdTe$_2$ based on its point group $D_{3d}$. Using the condition of a full gap left only three candidates remaining: an $s$-wave superconductor that transforms trivially under all lattice symmetries ($A_{1g}$), an odd-parity $p$-wave triplet state $(A_{1u})$ and a generically anisotropic triplet state $e_{u(1,0)}$. While the $s$-wave state is topologically trivial, the two odd-parity superconducting phases can exhibit non-trivial topology depending on the relative sign of the gap on the different Fermi pockets. 

One well-known approach to obtain further insight into the pairing symmetry is to investigate the behavior of the superconducting phase under tuning the amount of disorder in the system. In particular, the rate at which the transition temperature $T_c$ changes with the disorder level can provide information about the pairing state. This technique was successfully applied to various superconducting materials, for example, to the cuprates, ruthenates and the iron-based superconductors~\cite{TcsuppressionYBCO, SRO, BaRu, CaK1144, TcSuppressINs, ProzorovKogan2011, Hoyer2014}. Here, we use electron irradiation to study the impact of non-magnetic disorder on the superconducting and the normal state in single-crystals of PdTe$_2$. Irradiation with relativistic electrons in the MeV energy range at low-temperatures (at about $20$~K) is known as the most clean and controllable way to create point defects, predominantly in the form of vacancies and interstitials~\cite{dines}. Employing a combination of transport and London penetration depth measurements using a tunnel diode resonator (TDR) technique~\cite{vandegrift,Prozorov2006}, we find that the superconducting state remains fully gapped after irradiation. We observe that the transition temperature $T_c$ is suppressed with increasing levels of disorder, yet the rate of suppression is found to be notably lower than predictions of the Abrikosov-Gorkov (AG) theory for magnetic impurity scattering in $s$-wave superconductors~\cite{AG}. 

To interpret these experimental observations, one notes that, since the early work of Anderson~\cite{AndTh1}, and Abrikosov and Gorkov~\cite{AndTh2, AndTh3}, it is known that the superconducting state can enjoy protection against certain forms of disorder that obey appropriate symmetries: the transition temperature $T_c$ of a single-band, $s$-wave superconductor with an isotropic gap is independent of the amount of non-magnetic, i.e., time-reversal symmetric (TRS), disorder. This phenomenon is commonly referred to as  ``Anderson theorem''. In contrast, $T_c$ is reduced by the presence of magnetic impurities, i.e., time-reversal anti-symmetric (TRA) disorder. For a single-band, isotropic $s$-wave superconductor, $T_c$ then follows the well-known AG law~\cite{AG}. This is different for anisotropic gap functions, e.g, with (anisotropic) $s$-, $p$- or $d$-wave symmetry, for which $T_c$ is sensitive to TRS disorder already \cite{Chanin-PR-1959, Markowitz-PR-1963, Hohenberg1964, BWpWave,Maekawa,GolubovMarzin,RadtkeScattering,Puchkaryov,KoganScattering}. The decrease of $T_c$ as a function of an increasing scattering rate off non-magnetic impurities is therefore often (yet sometimes wrongly) regarded as a signature of unconventional superconductivity. 

The situation in multi-orbital and multi-band unconventional superconductors is significantly more rich~\cite{GolubovMarzin}. For example, the gap function can take different values~\cite{Wilke-PRB-2006} or even different signs~\cite{Mazin-Physica-2009, Bang-JPhysC-2017}, on different Fermi pockets, leading to a different sensitivity with respect to inter- and intraband scattering processes \cite{SpmScattering,TheorySpm,Wang2013}. 
Furthermore, spin-orbit coupling has been demonstrated to be able to enhance the stability of the superconducting state against disorder in both centrosymmetric~\cite{DisorderSOCFu,BrydonScattering} and non-centrosymmetric~\cite{OurDisorderSOC} multi-orbital systems.
Interestingly, a generalization of the Anderson theorem for the multi-orbital and multi-band case has recently been derived~\cite{Scheurer2016,Hoyer2015}, which shows that unconventional pairing states 
can also enjoy protection against certain forms of disorder. For instance, a two-band superconductor with a sign-changing $s^{+-}$ gap function is protected against TRA interband scattering as long as the size of the gap is equal on both Fermi surfaces~\cite{Hoyer2015}. 

In \refcite{Scheurer2016}, the general form of this Anderson theorem was derived and expressed in terms of (anti)commutators of the superconducting order parameter, the disorder potential, and the normal state Hamiltonian, thus, assuming a purely algebraic form that can be readily applied in any basis. We will review this form of the generalized Anderson theorem; we show that the rate at which $T_c$ decreases with increasing scattering strength is determined by a Fermi-surface average of precisely the same (anti)commutator that also enters the generalized Anderson theorem.
As a result, if the (anti)commutator-relations are only weakly violated, $T_c$ decreases slowly and superconductivity is significantly more protected than described by the AG law. We will see how special cases of the expression for the reduction of $T_c$ reproduce well-known results of the literature. 

The presence of strong spin-orbit coupling can largely suppress the rate at which $T_c$ decreases with disorder~\cite{DisorderSOCFu,OurDisorderSOC,Scheurer2016,BrydonScattering}. This results from a mixing of spin and orbital degrees of freedom that can potentially lead to a reduced overlap of the wavefunctions of scattering partners under the natural assumption that impurity scattering acts trivial in orbital space. Michaeli and Fu have shown in a $\vec{k}\cdot\vec{p}$ model relevant to doped Bi$_2$Se$_3$ that if the normal state Hamiltonian obeys an additional symmetry, such ``spin-orbit locking'' can even lead to a complete protection against disorder for fully gapped odd-parity superconductors~\cite{DisorderSOCFu}. We will see below that this result readily follows from the generalized Anderson theorem of \refcite{Scheurer2016}, revealing the general conditions for symmetry-enhanced protection of superconductivity against disorder.

Based on these insights and since the Fermi surface in PdTe$_2$ consists of several electron and two hole pockets, we analyze our experimental result of weak $T_c$ suppression under irradiation within the framework of this generalized Anderson theorem. This allows us to describe the different pairing scenarios within one framework. We employ the experimentally measured slope of the $T_c$ suppression with increasing scattering to make quantitative predictions on the properties of the different pairing state candidates. For instance, an $s$-wave pairing state that has the same sign on all Fermi surfaces, denoted by $A_{1g}^{++}$ below, must exhibit a rather substantial degree of momentum dependence of the superconducting gap to be consistent with the data. More precisely, the ratio of gaps on different Fermi sheets must be at least 2. 
Finally, the odd-parity $A_{1u}$ pairing is only consistent with the data, if the bands that make up the Fermi surface exhibit a substantial mixing of even and odd parity wavefunctions. This work exemplifies the predictive power of this combined experimental-theoretical approach to constrain the microscopic superconducting order parameter by controllably tuning the amount of non-magnetic disorder. At the same time, it also highlights important caveats in the interpretation of disorder-induced suppression of $T_c$ in multi-orbital systems with strong spin-orbit coupling.

The remainder of this paper is organized as follows. In \secref{sec:experimental_details}, we describe the experimental details of our measurement setup. Our experimental results of resistivity, Hall measurements, thermal transport, and London penetration depth before and after electron irradiation are presented in \secref{sec:results}. We discuss and interpret these results in \secref{sec:discussion} within the various possible superconducting pairing symmetries of the system. This allows us to draw quantitative conclusions, e.g., on the required degree of the superconducting gap anisotropy, and impose quantitative restrictions on the properties of the different pairing states based on our experimental results. We conclude in Sec.~\ref{sec:conclusions} and present details of the theoretical derivations and first-principle density functional theory (DFT) calculations in the Appendices. 

\section{Experimental details}
\label{sec:experimental_details}
Single crystals of PdTe$_2$ were grown using a procedure described in our earlier work~\cite{PRB}. Samples used for four-probe in-plane electrical resistivity, $\rho$, and thermal conductivity, $\kappa$, measurements and for 5-probe Hall effect measurements were cleaved from the inner parts of large single crystals with typical dimensions of (2-3)$\times$0.5$\times$0.05 mm$^3$. The longer side of the sample was along an arbitrary direction in the hexagonal crystal plane. Contacts to the fresh cleaved surface of the samples were made by attaching 50 $\mu$m silver wires with In solder  \cite{FeSedetwinning}.  The same samples were used before and after irradiation thus essentially eliminating the relatively large uncertainty associated with determining the geometric factor. The resistivity of the pristine samples at room temperature was set at 24 $\mu \Omega$cm as determined by statistically significant average on a large array of crystals in our previous study \cite{PRB}. Temperature dependent electrical resistivity and thermal conductivity measurements were made in two setups, PPMS (1.8 to 300~K) and cryogen free Janis $^3$He system (0.5 to 3~K). Modular thermal conductivity device was used \cite{MTCRSI}, enabling measurements in both systems without dismounting sample. For the Hall effect measurements the sample contacts were soldered to the side surfaces of 67 $\mu$m thick sample. Measurements were performed in PPMS device using magnetic field sweeps in the range $-9$T to $9$T at selected constant temperatures. The Hall resistance was determined as the difference between measurements in inverted magnetic fields.

We performed precision measurements of the in-plane London penetration depth $\Delta\lambda(T)$ using the tunnel-diode resonator (TDR) technique~\cite{vandegrift}. Measurements were conducted in a high stability $^3$He-cryostat with base temperature of $\sim$0.4~K. One sample was measured multiple times before and after electron irradiation. The sample was placed with its $c$-axis parallel to an excitation field, $H_{ac} \sim 20$ mOe, which is much smaller than $H_{c1}$~\cite{type1SC}. The shift of the resonance frequency, $\Delta f(T)=-G4\pi\chi(T)$, is proportional to the differential magnetic susceptibility $\chi(T)$. The constant $G=f_0V_s/2V_c(1-N)$ depends on the demagnetization factor $N$, sample volume $V_s$ and coil volume $V_c$. The constant $G$ was determined experimentally from the full frequency change that occurs when the sample is physically pulled out of the coil. To obtain the (change of the) London penetration depth $\Delta \lambda(T)$ as a function of temperature, we use the following expression $4\pi\chi=(\lambda/R)\tanh (R/\lambda)-1$~\cite{Prozorov2000,Prozorov2006}. Here, $R$ is an effective sample size that can can be calculated ad depends on the sample geometry, and $\chi(T)$ is the experimentally measured magnetic susceptibility.

Electron irradiation was performed at SIRIUS Pelletron linear accelerator in Laboratoire des Solides Irradi\'es at \'Ecole Polytechnique in Palaiseau, France. Relativistic  electrons with energy of 2.5 MeV were used to create point like defects (Frenkel pairs) by knocking the ions away from the regular position in the lattice \cite{Mizukami, SUST}. Details regarding electron irradiation and its influence on Fe-based superconductors can be found elsewhere \cite{SUST}. The defect concentration produced by irradiation with electrons with energies in the MeV range is homogeneous throughout the sample thickness as long as it is smaller than the relatively large electron penetration depth ($\sim$100 $\mu$m) \cite{Thompson}. The homogeneous damage of our samples can be seen directly from the fact that the superconducting transitions remain sharp after irradiation. The acquired irradiation dose presented in this paper is in the units of Coulomb per square centimeter, where 1 C/cm$^2$ = 6.24 $\times$ 10$^{18}$ electrons/cm$^2$. The total charge of electrons penetrated through the sample was measured by a Faraday cage placed behind the sample stage.

Contacts to the samples for transport measurements deteriorate after irradiation, resulting in a higher noise level. For this reason all transport measurements with different irradiation doses (0.91, 1.75 and 2.41 C/cm$^2$ for resistivity, 0.91 C/cm$^2$ for thermal conductivity  and 1.33 C/cm$^2$ for Hall effect) were made on individual samples, comparing pristine (before irradiation) and irradiated states. The $T_c$ for the sample with irradiation dose of 1.75 C/cm$^2$ was not determined. Multiple irradiation cycles allowing for accumulation of notably higher doses were used for samples used in penetration depth measurements, invoking no contact making. 

\section{Experimental results}
\label{sec:results}
\subsection{Electrical resistivity}
\label{subsec:electrical_resistivity}
We have measured longitudinal and Hall resistivity as a function of temperature and magnetic field, both before and after irradiation.
\begin{figure}
	\includegraphics[width=\linewidth]{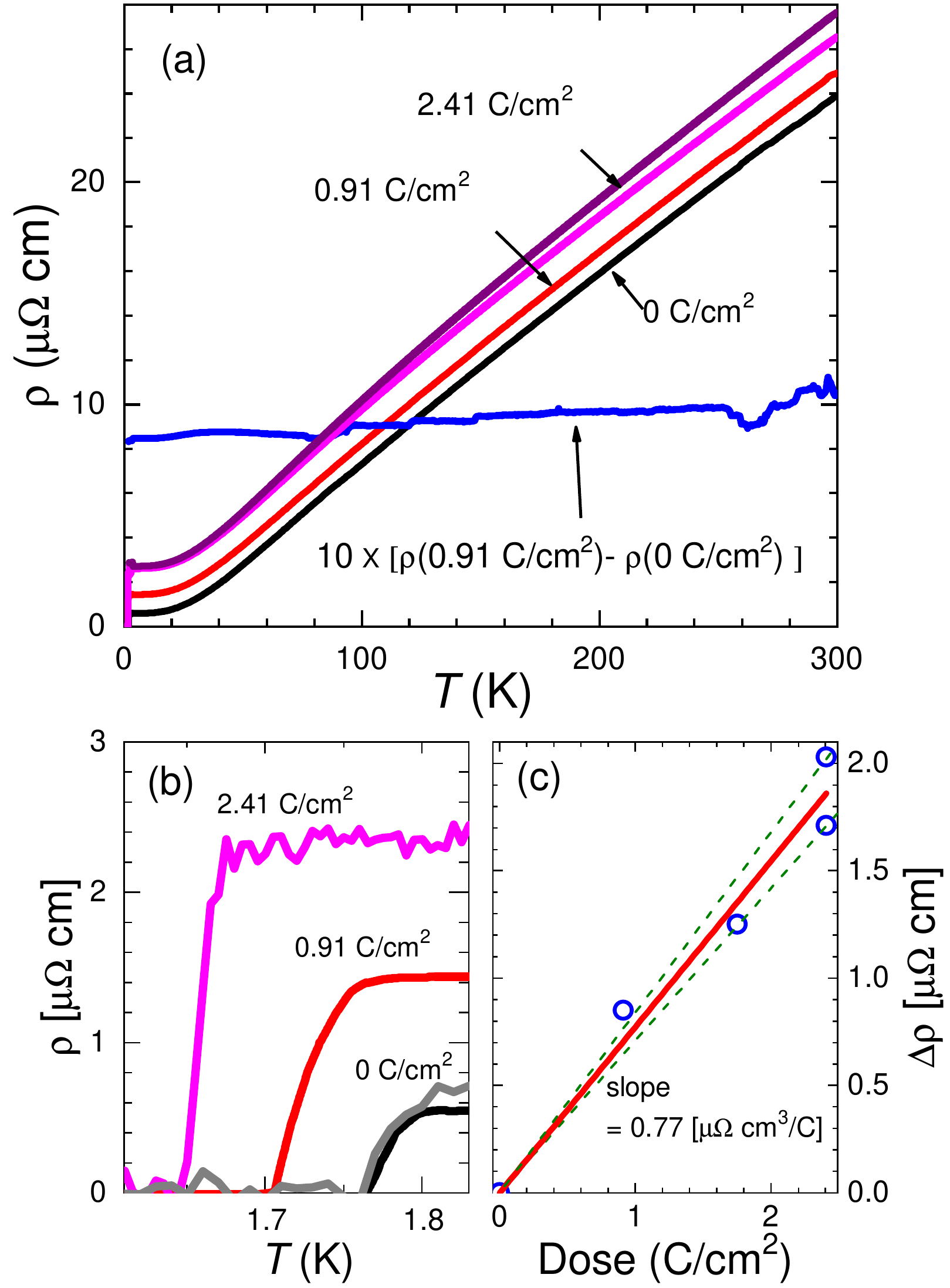}%
	\caption{(Color online)  (a) Temperature-dependent in-plane electrical resistivity of PdTe$_2$ before (black and grey curves) and after  low-temperature electron irradiations of 0.91 (red curve) and 2.41 C/cm$^2$ (magenta curves). All curves show a range of residual resistivity below approximately 10~K (in  which resistivity is approximately   temperature-independent) and linear increase above 40~K. The shift due to irradiation is almost  parallel, as can be seen in the temperature-independent difference between resistivities of the pristine and 0.91 C/cm$^2$ irradiated samples (multiplied by a factor of 10, the blue curve). The data for 2.41 C/cm$^2$ sample after irradiation do not show parallel $\rho(T)$ curve shift, as found in 0.91 C/cm$^2$ sample, indicating change of the geometric factor due to formation of cracks. Correction using normalized $\rho(T)$ curve slope near room temperature makes the 2.41 C/cm$^2$ look as a smooth parallel up-shift of  0.91 C/cm$^2$, light magenta curve. (b) The zoom of the superconducting transition range with $T_c$ suppression by 0.060~K and $\sim$0.11~K using resistivity offset criterion and residual resistivity increase from 0.6 to 2.3 $\mu \Omega$cm, from pristine to 2.41 C/cm$^2$. (c) The resistivity increase as a function of irradiation dosage. For sample with 2.41 C/cm$^2$ dose we use resistivity above $T_c$ without (top point) and with (bottom point) geometrical factor correction. The straight red line shows linear fit through all data point with slope of 0.77$\pm$0.07 $\mu \Omega$ per C/cm$^2$, the green dashed lines show the slopes for the error bar ranges. }
		\label{resistivity}
\end{figure}
The main panel of Fig.~\ref{resistivity} shows the temperature dependent in-plane resistivity $\rho$ of PdTe$_2$ before (black and grey) irradiation and after irradiations with the doses of 0.91 C/cm$^2$ (red curve) and 2.41 C/cm$^2$ (magenta curves). The observed temperature dependence is typical of a good metal with a range of nearly temperature independent resistivity below 10~K and a linear increase of $\rho$ with $T$ above 40~K. The response of the sample resistivity to disorder introduced by electron irradiation is also typical of a simple metal. As expected from Matthiessen rule, the curves for 0.9 C/cm$^2$ irradiation shift up parallel to themselves due to an increase of residual resistivity from $\rho^{(\text{pristine})}(0) = 0.6 \mu\Omega$cm  to $\rho(0)^{(\text{2.41 C/cm$^2$})}$=2.3 $\mu \Omega$cm, see panel (b). The resistivity difference of the irradiated and pristine samples, 
$\rho(0.91~\rm{C/cm^2}) - \rho(0~\rm{ C/cm^2})$, as shown with blue line in the main panel (a) of Fig.~\ref{resistivity} magnified by a factor of 10, is almost temperature independent.
For sample with 2.41 C/cm$^2$ irradiation  (dark magenta line) the shift is not parallel, and the slope of the line above 40~K increases. This observation suggests that the geometric factor of the sample changed during irradiation due to crack formation \cite{SUST}. Partially this effect can be removed by normalizing the slope of the curve at high temperature to that before irradiation (light magenta curve). This brings approximately 10\% uncertainty to the residual resistivity of the 2.41 C/cm$^2$ irradiated sample. In panel (c) we show increase of residual resistivity with irradiation dose. For 2.41 C/cm$^2$ sample we show two values as determined from direct measurements (top point) and from the slope-normalized curve (bottom point). The dependence of $\rho(0)$ on dose is close to linear, as expected and observed in samples continuously measured {\it in-situ} at low temperatures \cite{ProzorovPRX}. Due to variation of the geometric factor, the slope of the curve is determined by linear fit through data points with onset fixed at (0,0) (red curve) as 0.77 $\mu\Omega$cm  per C/cm$^2$. Green dashes show error bars of slope determination.  We use linear dependence of residual resistivity on dose with the
slope of 0.77 $\mu \Omega$cm/(C/cm$^2$) for determination of resistivity in samples used in penetration depth studies, see Fig.~\ref{resistivity}(c) below. 
Importantly, the superconducting transition is equally sharp before and after irradiation treatments, with a full width of just 0.05~K from the onset to zero resistivity. The transition temperature $T_c$ is suppressed by $0.06$~K as a result of irradiation with a dose of 0.91 C/cm$^2$ and by $0.11$~K for a dose of 2.41~C/cm$^2$. 

\begin{figure}[t!]
	\includegraphics[width=\linewidth]{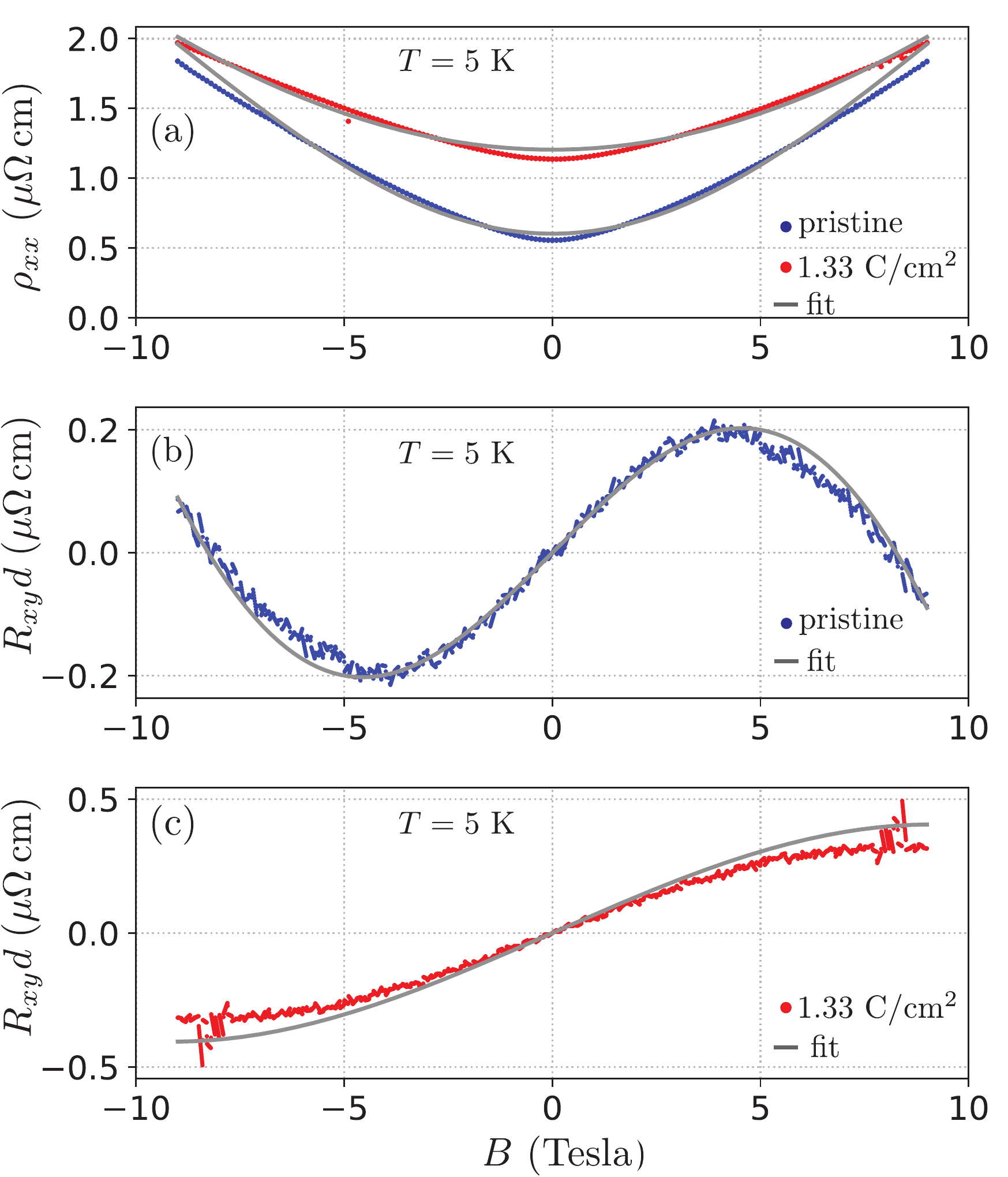}\\%
	\caption{(Color online) (a) Low-temperature longitudinal resistivity $\rho_{xx}$ as a function of magnetic field $B = \mu_0 H$ at $T=5$~K. Blue (red) line shows result for the pristine (irradiated with dose $1.33$~C/cm$^2$) and grey line is two-band model fit. (b-c) Low-temperature Hall resistance $R_{xy}$ multiplied by sample thickness $d=67 \mu$m as a function of magnetic field $B$ of pristine sample (blue, panel b), irradiated sample (red, panel c) together with two-band model fit (grey). 
	}
	\label{Hall}
\end{figure}

% Hall data
In Fig.~\ref{Hall}, we show the longitudinal resistivity $\rho_{xx}$ and transverse Hall resistance multiplied by the sample thickness $d$, $R_{xy}d$, as a function of magnetic field at low temperatures $T=5$~K, both before and after electron irradiation with a dose of $1.33$~C/cm$^2$. To extract the carrier densities and mobilities, we perform a fit to standard expressions of $\rho_{xx}$ and $\rho_{xy}$ for a two-band model of electron and hole charge carriers (for details see Appendix~\ref{sec_app:two_band_model_fit_resistivity})~\cite{Pippard}. For the pristine sample, the optimal fit parameters are $n^{(\text{pristine})}_e = 4.2(1) \times 10^{27}$m$^{-3}$, $n^{(\text{pristine})}_h = 2.2(1) \times 10^{27}$m$^{-3}$ for the electron and hole densities, and $\mu^{\text{(pristine)}}_e= 0.10(1) \text{m}^2/\text{Vs}$ and $\mu^{\text{(pristine)}}_h = 0.28(1)\text{m}^2/\text{Vs}$ for the electron and hole mobilities. After irradiation with a dose of $1.33$~C/cm$^2$, the densities $n_e$ and $n_h$ are unchanged, and the mobilities are reduced by approximately a factor of two: $\mu^{\text{(1.33 C/cm$^2$)}}_e= 0.05(1)\text{m}^2/\text{Vs}$ and $\mu^{\text{(1.33 C/cm$^2$)}}_h = 0.14(1)\text{m}^2/\text{Vs}$. This is consistent with our observation that the Hall constant $R_H$ at low temperatures is unchanged during irradiation with this dose. We note that $R_H = 0.60(5) \text{mm}^3/\text{C}$ is approximately independent of temperature in the prisitine sample. 

Using the bare electron mass, we can extract a rough estimate of the scattering rates $h/\tau_e = h e_0/(m_e \mu_e) = 70(2)$~meV and $h/\tau_h = 26(2)$~meV of the pristine sample. The rates are a factor of two larger after irradiation. Using the approximation of a three-dimensional quadratic dispersion, we find mean-free paths of $\ell_e = 172(5)$~nm and $\ell_h = 363(5)$~nm for electron and hole charge carriers after irradiation (with dose $1.33$ C/cm$^2$). Note that this is of the same order as the superconducting coherence length $\xi = 439$~nm, reported in the material~\cite{type1SC, PRB}. Importantly, this corresponds to a small disorder parameter $k_F \ell_e = 860(5)$ and $k_F \ell_h = 145(5)$ for electrons and holes, respectively, justifying the perturbative treatment of disorder we use below.

\begin{figure}[tbh]
    \includegraphics[width=0.90\linewidth]{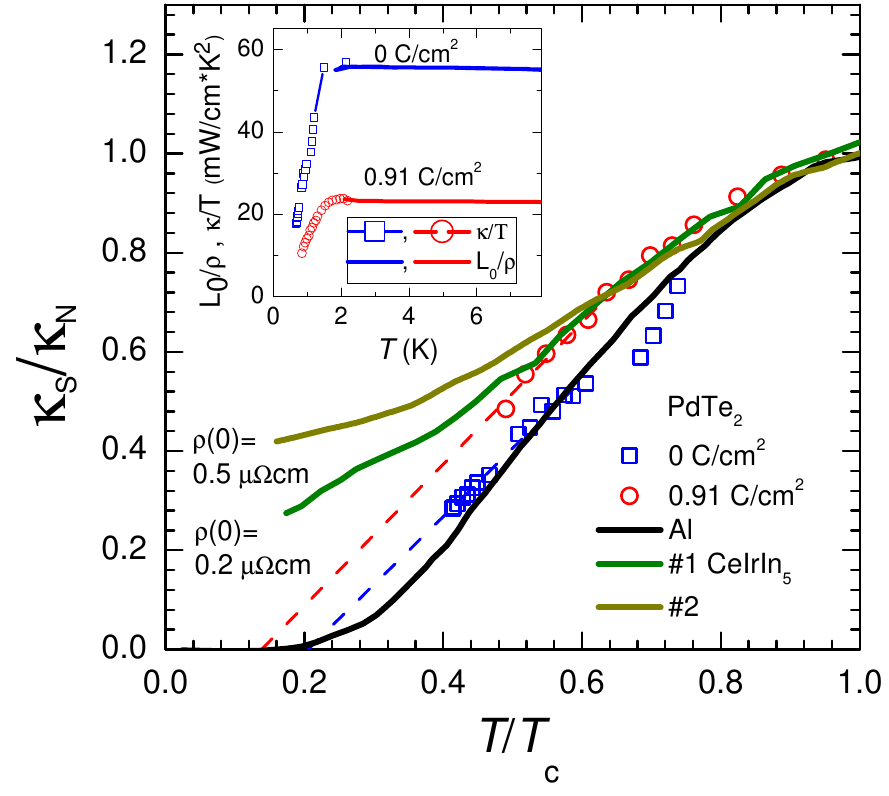}
	\caption{Inset shows comparison of thermal conductivity, $\kappa/T$, (symbols) and electrical resistivity data converted to equivalent thermal conductivity by using the WF law, $L_0/\rho$ (lines). The two data sets coincide within accuracy of our measurements above $T_c$ in both pristine (blue) and irradiated (red) samples, signifying the validity of WF law. Main panel shows ratio of thermal conductivity in the superconducting and normal states,  $\kappa_S(T)$/$\kappa_N(T_c)$, plotted versus reduced temperature $T/T_c$. Open blue squares and red circles are the data for pristine and electron-irradiated PdTe$_2$ with dose $0.91$~C/cm$^2$.  The blue and red dashed lines are linear fit to the low temperature part of the data, showing extrapolation towards  a negative value, suggesting a fully gapped superconducting state. For comparison we show data for the conventional full-gap superconductor Al~\cite{Al} extrapolating to a negative value, and the nodal superconductor CeIrIn$_5$~\cite{Hamidehhybrid} for samples with different quality as characterized by the residual normal state resistivity of $\sim$0.2 $\mu \Omega$cm (green line) and $\sim$0.5 $\mu \Omega$cm (dark yellow line), with linear extrapolation to a finite ratio as $T\rightarrow 0$.
    }
	\label{kappakappaN}
\end{figure}

\subsection{Thermal transport}
We have measured the thermal conductivity $\kappa$ in PdTe$_2$ before and after irradiation with a dose of $0.91$~C/cm$^2$. Measurements of $\kappa$ were made in  a temperature range from $T=0.5$~K to $T=3$~K using MTC unit in cryogen free He3 setup (see Sec.~\ref{sec:experimental_details}). In the normal state, above $T_c$ we observe that $\kappa$ is related to the electrical conductivity $\sigma = 1/\rho$ via the Wiedemann-Franz (WF) law $\kappa/\sigma = L_0 T $ within experimental scatter of the data to the accuracy of better than 10\%, see the inset in Fig.~\ref{kappakappaN}.   Here, $L_0=\frac{\pi^2}{3}(\frac{k_B}{e})^2=$2.45$\times 10^{-8}\text{W}\Omega \text{K}^{-2}$ is the Sommerfeld value of the Lorenz number ~\cite{WF,NJP,DasSarma}. This shows that the phonon contribution to the thermal conductivity is negligible in the normal state. The WF law is obeyed in both the pristine and the irradiated sample. 

In the superconducting state, the WF law is grossly violated, because the superfluid part of the conduction electron density does not contribute to thermal conductivity. The electronic part of thermal conductivity is determined by the residual electronic excitations, which in the case of a fully gapped superconductor are exponentially suppressed as $T \to0$~\cite{NJP}. Therefore, a straightforward way to distinguish between nodal and full-gap superconductors is to extrapolate $\kappa(T)$ to its value as $T \to 0$. For a nodal superconductor, this extrapolation yields a finite positive value~\cite{SigristUeda}, while $\kappa/\kappa_N$ extrapolates to a (physically meaningless) negative value in full-gap case. The contribution of phonons in more disordered samples is more significant, which makes this extrapolation to negative value less convincing in the sample subjected to 0.91 C/cm$^2$ electron irradiation.

In Fig.~\ref{kappakappaN}, we plot the thermal conductivity normalized by its value at $T_c$, $\kappa_S(T)/\kappa_N(T_c)$, which shows that $\kappa$ extrapolates to a negative value, corresponding to a full gap, for both pristine and irradiated samples of PdTe$_2$. For comparison, we plot the thermal conductivities of Al~\cite{Al}, a clean fully gapped superconductor, which shows negative $\kappa_S(T)/\kappa_N(T_c)$ extrapolation as $T\to 0$, and the nodal superconductor CeIrIn$_5$~\cite{Hamidehhybrid,Hamidehuniversal} for samples of different disorder level. In the nodal case, the curve  extrapolates to a finite positive value~\cite{Tailleferuniversal,Suzukiuniversal,Hamidehuniversal} rapidly increasing for samples with large residual resistivity~\cite{Hamidehuniversal,Suzukiuniversal}, compare curves for samples with $\rho (0) \sim 0.2$ and 0.5 $\mu \Omega$cm.

\subsection{London penetration depth}
\label{subsec:London_penetration_depth}
\begin{figure}[tb]
\begin{center}
\includegraphics[width=0.90\linewidth]{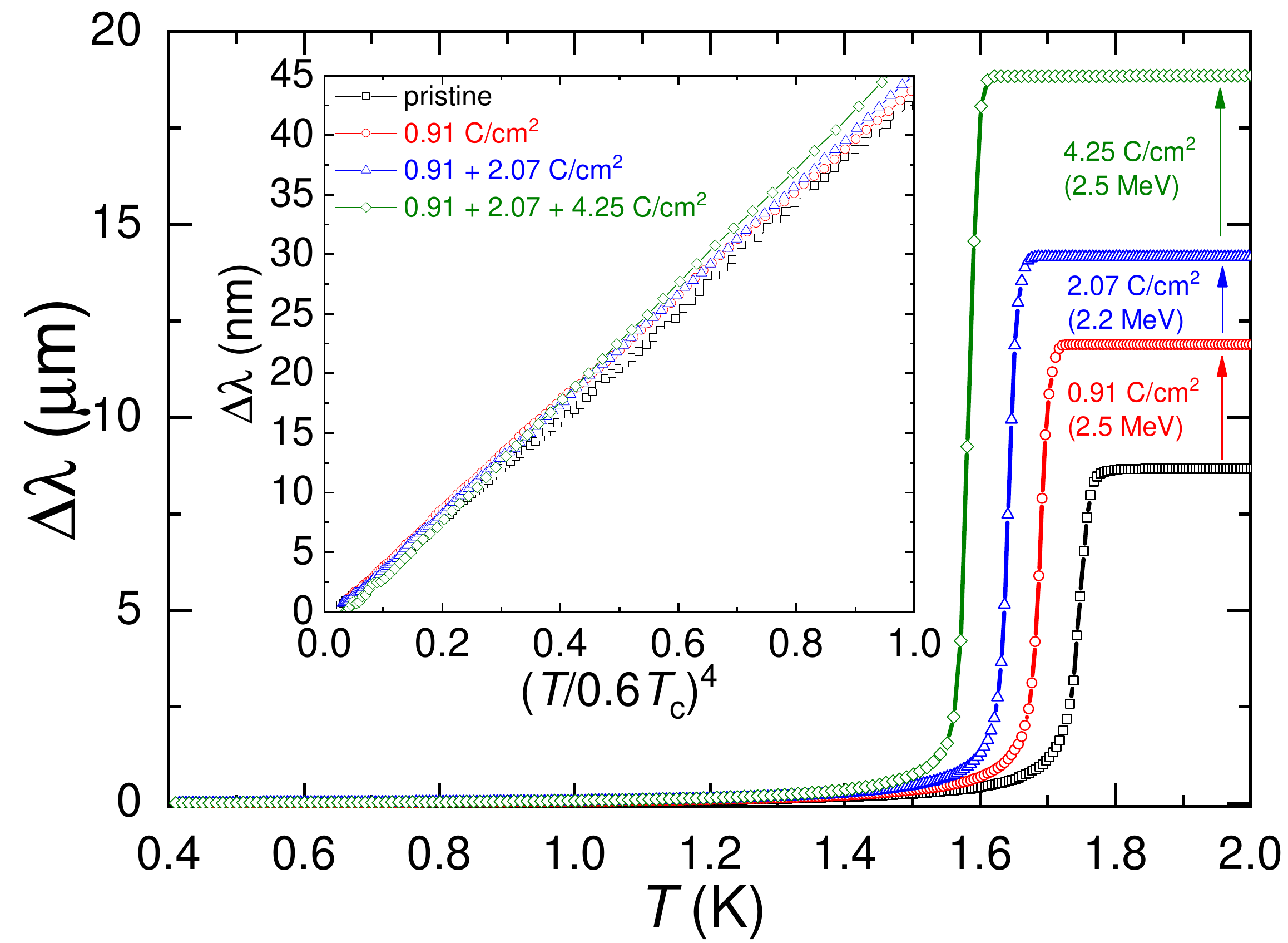}
\end{center}
\caption{(color online) Temperature variation of London penetration depth $\Delta \lambda (T)$ measured in $^3$He TDR setup before and after irradiations. Main panel displays data over the whole temperature range that clearly shows the suppression of $T_c$ and increase of normal state skin depth ($T$ $>$ $T_c$). Inset shows low temperature penetration depth below 0.6 $T_c$ plotted as a function of $T^4$, which reveals the exponential behavior of BCS-type full gap structure even after three irradiations.
}%
\label{DeltaLambda}
\end{figure}

\begin{figure}[tb]
\begin{center}
\includegraphics[width=0.90\linewidth]{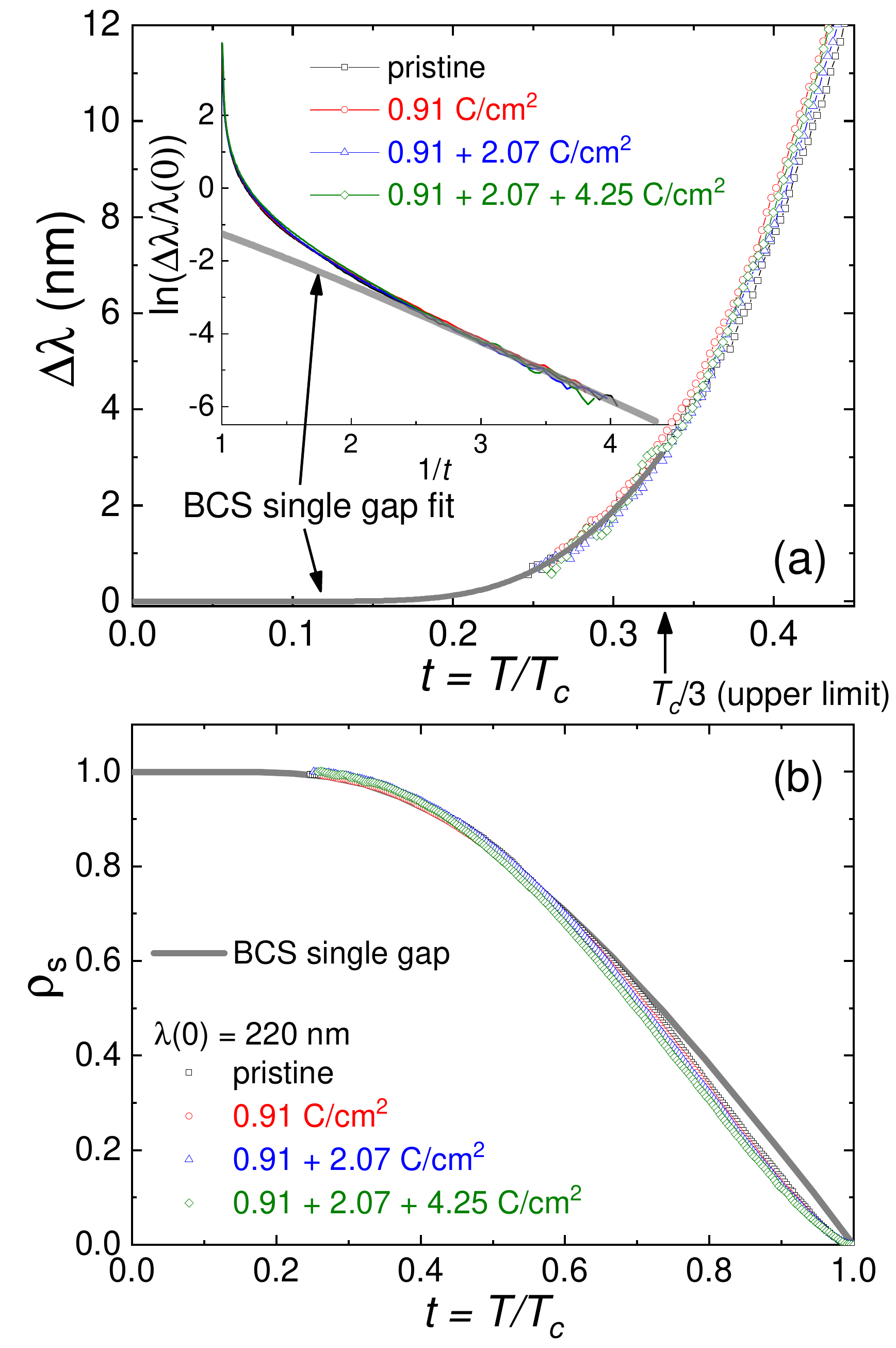}
\end{center}
\caption{(Color online) (a) The low-temperature region of London penetration depth before and after irradiations (identical sample in Fig.~\ref{DeltaLambda}). All data are well-fitted with BCS single gap fuction. The inset show $\Delta \lambda \propto (T/T_c)^4$ over the range up to $0.6T_c$ in both the pristine state and after two irradiations. (b) Superfluid density calculated from $\Delta \lambda$ in panel (a), $\rho_s(T)\equiv (\lambda(0)/\lambda(T))^2$,  which is found to be only very weakly affected by irradiation. %extremely robust against irradiation.
}%
\label{rhos}
\end{figure}

In the main panel of Fig.~\ref{DeltaLambda}, we present the temperature dependent London penetration depth $\Delta \lambda$, measured over the whole range of superconductivity. In the pristine sample, the superconducting transition occurs at temperature $T_c=1.76$~K, where we have used $\rho(T_c) = 0$ as the criterion to define $T_c$. The transition is sharp and highly reproducible, as expected in stoichiometric materials. Upon a series of electron irradiation with doses of 0.91 C/cm$^2$ (2.5 MeV), plus 2.07 C/cm$^2$ (2.2 MeV) and plus 4.25 C/cm$^2$ (2.5 MeV), the transition temperature continuously decreases from 1.76 to 1.70 K (for a dose of 0.91 C/cm$^2$) to 1.66 K (for an additional dose of 2.07 C/cm$^2$) to 1.59 K (for additional dose of 4.25 C/cm$^2$). This corresponds to a decrease by about 9.6\% in total. 
The increase of the penetration height above $T > T_c$ upon irradiation is caused by an increase of the normal state skin depth and is consistent with the observed increase of residual resistivity $\rho(0)$ from a direct measurement in Fig.~\ref{resistivity}. 

The inset of Fig.~\ref{DeltaLambda} shows $\Delta \lambda(T)$ at low-temperatures before and after irradiation. We see that $\Delta \lambda \propto (T/T_c)^4$ over the range up to $0.6T_c$ in both the pristine state and after two irradiations. It is quite remarkable that despite the notable $T_c$ suppression, the functional dependence of $\Delta \lambda$ on the temperature is almost unchanged at low temperatures and remains $\propto T^4$. In the temperature range that we consider, a power-law function with exponent $n \approx 4$ cannot be distinguished from an exponential function. An exponential decrease of  $\Delta \lambda(T)$ is expected in a clean and fully gapped BCS superconductors~\cite{BCS}. In contrast, in the presence of line nodes in the gap, one rather expects a close to $T$-linear decay of $\Delta \lambda$. In both cases, the behavior changes to $T^2$ under the addition of sufficiently strong disorder~\cite{ProzorovKogan2011}. Note that our study is not performed within this strong disorder regime and $\Delta \lambda$ remains $\propto T^4$ for all irradiation doses. 

It is important to restrict the fit of $\Delta \lambda$ to a low-temperature region below $0.33T_c$, where the temperature dependence of the superconducting gap magnitude is negligible (in single gap superconductors) and the $T$-dependence is determined by thermal excitation of quasi-particles across the superconducting gap. In the top panel of Fig.~\ref{rhos}, we show a fit of the penetration depth $\Delta \lambda$ using an exponential temperature dependence, which is expected from Bardeen-Coooper-Schrieffer (BCS) theory. We obtain an excellent fit using the single gap isotropic BCS expression
\begin{equation}
   \Delta \lambda(T) = \lambda(0) \sqrt{\frac{\pi \delta}{2 t}} \exp(-\delta/t)
\end{equation}
with $\delta$ = $\Delta$/$T_c$ = 1.76 and $t = T/T_c$. The fit yields a zero temperature value of $\lambda$(0) = 220$\pm$15~nm (pristine), 235$\pm$15~nm (0.91 C/cm$^2$), and 214$\pm$15~nm (0.91 + 2.07 C/cm$^2$). The data after 0.91 + 2.07 + 4.25 C/cm$^2$ show higher noise level preventing us from a good fitting to get $\lambda$(0). In the inset of Fig.~\ref{rhos} (a), we also plot $\mathrm{ln}(\Delta\lambda/\lambda(0))$ versus 1/$t$, clearly showing that all data are consistent with BCS single gap fit. We conclude that the zero temperature penetration depth $\lambda(0)$ is approximately constant within error bars. This is in reasonable agreement with expectations based on $\lambda_{\text{eff}}(\ell) = \lambda(0) ( 1 + \xi/\ell)^{1/2}$~\cite{Tinkham}, and using that the carrier densities are unchanged by irradiation (and $\lambda(0)$ thus remains unchanged) and $\ell_e, \ell_h \approx \xi$. Our estimated mean-free path $\bar{\ell}^{\text{(pristine)}} \approx 530$~nm (average of $\ell_e$ and $\ell_h$) is of the same order as the superconducting coherence length $\xi =439$~nm~\cite{PRB, type1SC}. We thus expect that $\lambda(0)^{(\text{pristine)}}/\lambda(0)^{(0.91 \text{C/cm}^2)} \approx 0.85$, while we find approximately $220/235 \approx 0.94$. In the following, we work with an average value of $\lambda(0) = 220$~nm.

Having experimentally determined $\lambda(0)$ using the BCS fit allows us to construct the temperature-dependent normalized superfluid density as $\rho_s(T)=(\lambda(0)/\lambda(T))^2$, with $\lambda(T)=\lambda(0)+\Delta \lambda (T)$. In the bottom panel of Fig.~\ref{rhos} we show the resulting superfluid density $\rho_s(T)$ before and after irradiation, which is calculated from our experimental data and using an average value of $\lambda(0)$ = 220~nm. The data are plotted versus reduced temperature $T/T_c$, and compared with BCS expectations for a single fully gapped superconductor (thick grey curve). The excellent agreement clearly shows that superconductivity in PdTe$_2$ can be well characterized by a \emph{single} and full superconducting gap energy scale. Importantly, this observation sets rather stringent conditions on the amount of anisotropy of the gap magnitudes $|\Delta_\alpha|/|\Delta_\beta|$ that are consistent with this behavior, despite the rather substantial $T_c$ suppression upon increasing disorder.

\section{Discussion}
\label{sec:discussion}

\begin{figure}
	\includegraphics[width=0.90\linewidth]{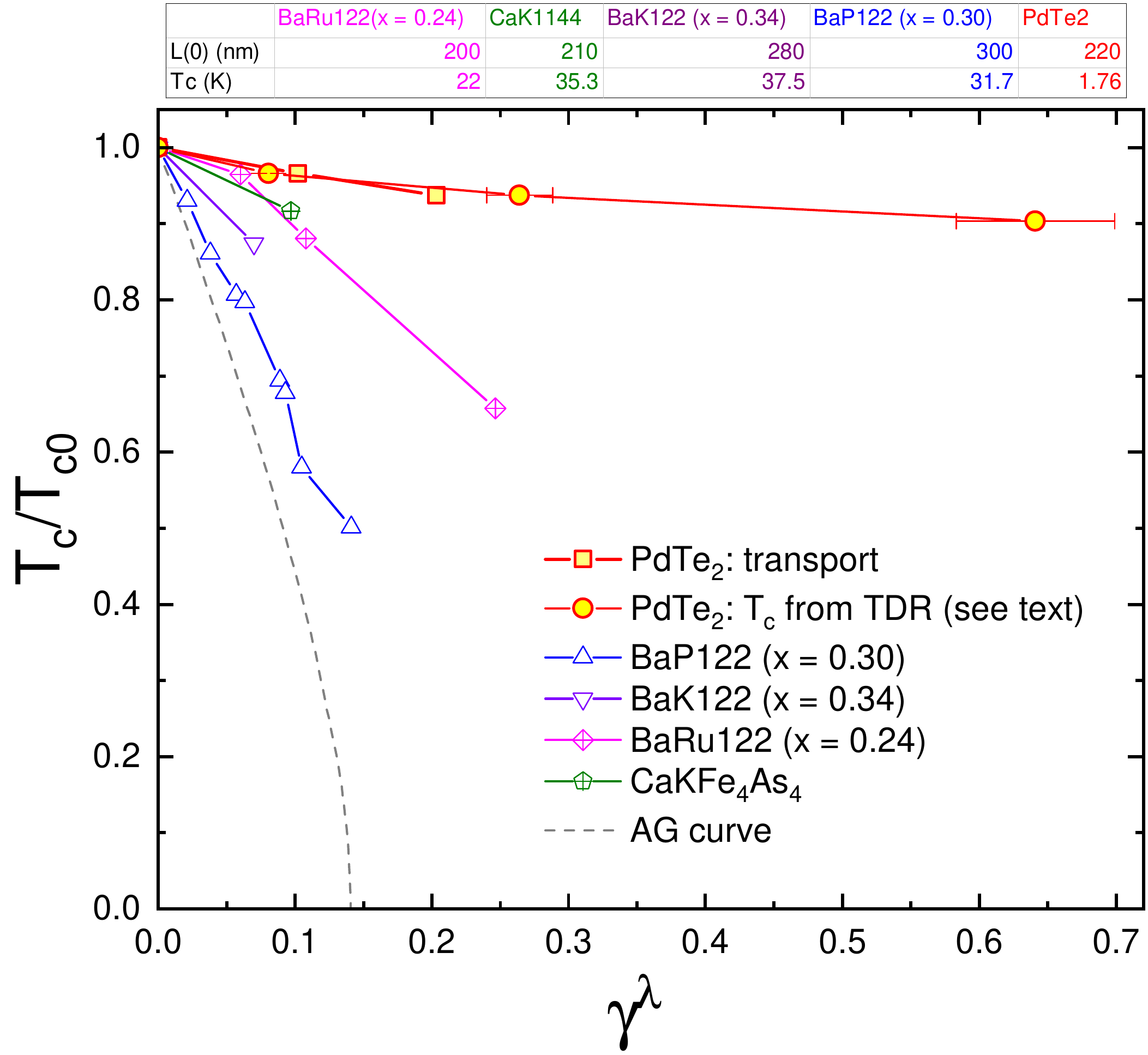}%
	\caption{(Color online)  $T_c$ suppression rate upon electron irradiation in PdTe$_2$ in comparison with several representative cases. The temperature axis is normalized by $T_{c0}$ in the pristine state and the scattering rate $\gamma^{\lambda}$ is calculated based on the increase in resistivity upon irradiations (See text). For comparison, reports on other materials are plotted together: BaRu122 (x = 0.24) \cite{ProzorovPRX}, CaK1144 \cite{CaK1144} , BaK122 (x = 0.34) \cite{KCHO2014PRB_BaK122}, and BaP122 (x = 0.3) \cite{Mizukami2017JPSJ_BaP122}.}
	\label{Tcgamma}
\end{figure}

In this section, we analyze our experimental results with the goal of determining properties of the superconducting pairing state, for example, its pairing symmetry and gap anisotropies. Our comprehensive analysis puts important quantitative constraints on the superconducting state in each of the possible full-gap pairing scenarios~\cite{PRB}. While we cannot definitely rule out any of the candidate states, our analysis points towards an unconventional spin-singlet $A_{1g}^{+-}$ pairing state as the most likely candidate. Further work addressing microscopic details of disorder scattering caused by electron irradiation is suggested. 

Let us briefly summarize the main experimental results presented in Sec.~\ref{sec:results}. We have found that electron irradition of PdTe$_2$ single crystals leads to a temperature independent upward shift of the resistivity $\rho$ that is caused by an increase in the residual resistivity $\Delta \rho_0$, which is proportional to the irradiation dose. Analyzing Hall resistivity demonstrates that irradiation leaves carrier densities unchanged and merely reduces mobility. This is consistent with an increase in the non-magnetic scattering rate $\tau^{-1}$ caused by the (expected) creation of point-like Frenkel-pair defects. The superconducting transition temperature $T_c$ decreases with increasing residual resistivity $\rho_0$. Specifically, it changes from $T_{c,0} = 1.76$~K in the pristine sample, where $\rho_0 = 0.6 \mu\Omega \text{cm}$, to $T_c(2.41 \text{C/cm$^2$}) = 1.65$~K, where $\rho_0 = 2.3 \mu \Omega \text{cm}$. London penetration depth $\lambda$ and thermal conductivity $\kappa$ measurements show that the superconducting state remains fully gapped. Importantly, the temperature dependence of $\lambda(T)$ can be well described by a \emph{single} gap energy scale. 

\subsection{Suppression of $T_c$ with irradiation dose}
\label{subsec:suppression_of_Tc}
A measurement of the rate of $T_c$ suppression with increasing levels of non-magnetic disorder can be used to distinguish different superconducting pairing states (see e.g.~\cite{ProzorovPRX, Wang2013, SUST, Hirschfeld_review}). To quantitatively analyze the suppression of $T_c$ with irradiation dose, we plot in Fig.~\ref{Tcgamma} the transition temperature ratio $T_c/T_{c,0}$ as a function of the experimentally determined dimensionless scattering parameter~\cite{ProzorovPRX} (see also Appendix~\ref{sec_app:dimensionless_scattering_rate})
\begin{equation}
    \gamma^\lambda = \frac{\hbar}{2 \pi k_B \mu_0 } \frac{\Delta \rho_0}{\lambda_0^2 
    T_{c,0}} = 0.98 \frac{\Delta \rho_0 [\mu \Omega \, \text{cm}]}{ \lambda_0^2 [10^{-7}\text{m}] T_{c,0} [\text{K}]} \,.
\end{equation}
Here, $\Delta \rho_0$ is the residual resistivity change due to electron irradiation and $\lambda_0$ is the zero temperature penetration depth. In Fig.~\ref{Tcgamma}, there are two different sets of PdTe$_2$ data: one from transport (Fig.~\ref{resistivity}) and the other from TDR (Fig.~\ref{DeltaLambda}), respectively. For TDR data, we do not have corresponding resistivity values. So, we used the slope in Fig.~\ref{resistivity} (c), and converted the dose to $\Delta \rho_0$. Then, this $\Delta \rho_0$ is used to calculate $\gamma^{\lambda}$. Both curves of PdTe$_2$ are consistent with each other.

We extract important information about the superconducting state from the observed (initial) slope of the $T_c$ suppression. The upper limit (i.e. fastest) suppression is the well-known AG law~\cite{AG}: $\delta T_c/T_{c,0} = - \frac{\pi^2}{2} \gamma^\lambda$ (dashed line in Fig.~\ref{Tcgamma}). This occurs when all scattering processes are pair breaking, which is, for example, the case for TRA scattering off magnetic impurities in an isotropic, single-band spin-singlet superconductor. The lower limit corresponds to the Anderson theorem, which is the case where $T_c$ is completely unaffected by scattering $\delta T_c/T_{c,0} = 0$~\cite{AndTh1, AndTh2, AndTh3}. This occurs, for example, for TRS scattering off non-magnetic impurities in the isotropic single-band $s$-wave case. The phenomenology becomes richer in the presence of spatial anisotropies of the gap and multiple orbitals or bands~\cite{Hohenberg1964,BWpWave,Maekawa,RadtkeScattering,GolubovMarzin,Puchkaryov,KoganScattering,Hoyer2015,Scheurer2016}. We have therefore included the experimentally observed $T_c$ suppression upon electron irradiation in various iron-based superconductors Ba(Fe$_{1-x}$Ru$_x$)$_2$As$_2$ (x = 0.24)
~\cite{ProzorovPRX}, CaKFe$_4$As$_4$~\cite{CaK1144}, (Ba$_{1-x}$K$_x$)Fe$_2$As$_2$ (x = 0.34)~\cite{KCHO2014PRB_BaK122}, and BaFe$_2$(As$_{1-x}$P$_x$)$_2$ (x = 0.3)~\cite{Mizukami2017JPSJ_BaP122}.

Below, we discuss in detail under which conditions the different potential pairing states in PdTe$_2$ could give rise to the observed finite suppression rate. Comparison with other experimental observations will then be used to rule out (or at least disfavor) certain states. For example, the fact that $\lambda(T)$ is well described by a single gap energy scale limits the degree of possible gap anisotropy in the system. 

To analyze the different pairing scenarios we make use of a generalized Anderson theorem that was recently derived by one of us~\cite{Scheurer2016}. The theorem is stated in terms of (anti)commutators of the superconducting order parameter, the disorder potential and the Hamiltonian in the normal state. We show below that the rate of $T_c$ suppression is determined by the value of these (anti)commutators, averaged over the Fermi surface. As a result, the suppression rate is small if the (anti)commutators are only weakly violated on average. 

In the following, we describe the possible superconducting pairing states and their topology, before deriving the conditions under which their respective superconducting $T_c$ would be suppressed with the experimentally observed rate. 
\subsection{Candidate pairing states}
As analyzed in detail in our previous work \cite{PRB}, the point group $D_{3d}$ and Fermi surface topology of PdTe$_2$ allows for only three distinct pairing symmetries with a full superconducting gap that are, hence, consistent with our penetration depth and thermal transport data: the topologically trivial $s$-wave, spin-singlet superconductor transforming under the irreducible representation (IR) $A_{1g}$ and the two $p$-wave triplet states, $A_{1u}$ and $E_u(1,0)$. The latter two triplet states \textit{can} be topologically nontrivial depending on microscopic details (see below). As the $E_u(1,0)$ state breaks rotational symmetry, its gap is in general anisotropic. As such, the exponential behavior of $\lambda(T)$ observed at all temperatures requires fine-tuning for the $E_u(1,0)$ state. Therefore, we will mostly focus on the $A_{1g}$ and $A_{1u}$ states in the following discussion. 

In any system, such as PdTe$_2$, with both time-reversal, $\Theta$, and inversion, $P$, symmetries, all bands have to be doubly-degenerate and one can introduce a \textit{pseudospin basis} at each $\vec{k}$-point with the same transformation properties as the electron's spin. As will become important below, the associated basis states are in general complicated, $\vec{k}$-dependent admixtures of spin and orbital degrees of freedom in a spin-orbit-coupled multi-orbital system such as PdTe$_2$. Denoting the Pauli matrices in pseudospin space by $\sigma_j$, $j=1,2,3$, the corresponding order parameters can be written as 
\begin{equation}
    \Delta_{\vec{k}} = \Delta_0(\vec{k}) i \sigma_2
\end{equation}
for the $A_{1g}$ state, and 
\begin{equation}
    \Delta_{\vec{k}} = \sum_{j=1}^3(\vec{d}_{\vec{k}})_j\sigma_j i\sigma_2
\end{equation}
with $\vec{d}_{\vec{k}} = d_0( X_{\vec{k}}, Y_{\vec{k}},Z_{\vec{k}})$ for the $A_{1u}$ state~\cite{PRB}. Here $\Delta_0(\vec{k})$ transforms as a scalar under all symmetry operations $g$ of the point group $D_{3d}$, $\Delta_0(g\vec{k})=\Delta_0(\vec{k})$, and $X_{\vec{k}}$, $Y_{\vec{k}}$, and $Z_{\vec{k}}$ transform as $k_x$, $k_y$, and $k_z$ under $D_{3d}$. Motivated by the experimental observation of a fully established gap, we assume (unless stated otherwise) absence of accidental nodes, $\Delta_0(\vec{k}),|\vec{d}_{\vec{k}}| \neq 0$ for $\vec{k}$ in the vicinity of a Fermi surface of the system. 

According to band structure calculations and experiment~\cite{dHvA,PdTe-ARPES}, the system has several Fermi surfaces. This is confirmed by our DFT calculations using a full-potential linear augmented plane wave (FP-LAPW) method, as implemented in \textsc{wien2k}~\cite{WIEN2k} (for details on DFT see Appendix~\ref{sec_app:DFT}). As shown in Fig.~\ref{fig:dft_main_text}, there are two hole pockets enclosing the $\Gamma$ point, both of which arise from bands associated with the type-II Dirac cone below the Fermi level, and additional Fermi surfaces around the K and K$'$ points. 
\begin{figure}
    \centering
    \includegraphics[width=.52\linewidth,clip]{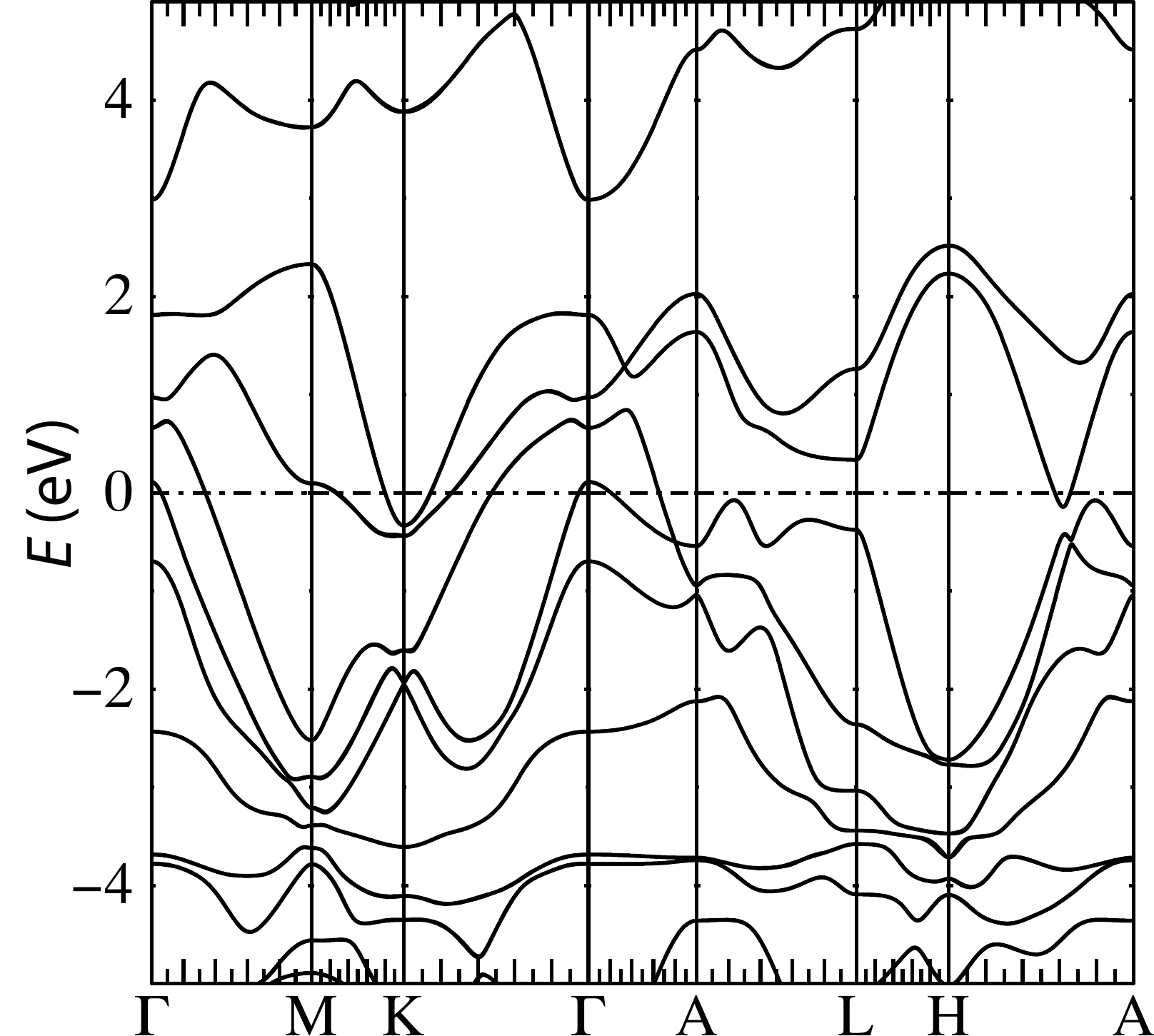}
    \includegraphics[width=.46\linewidth,clip]{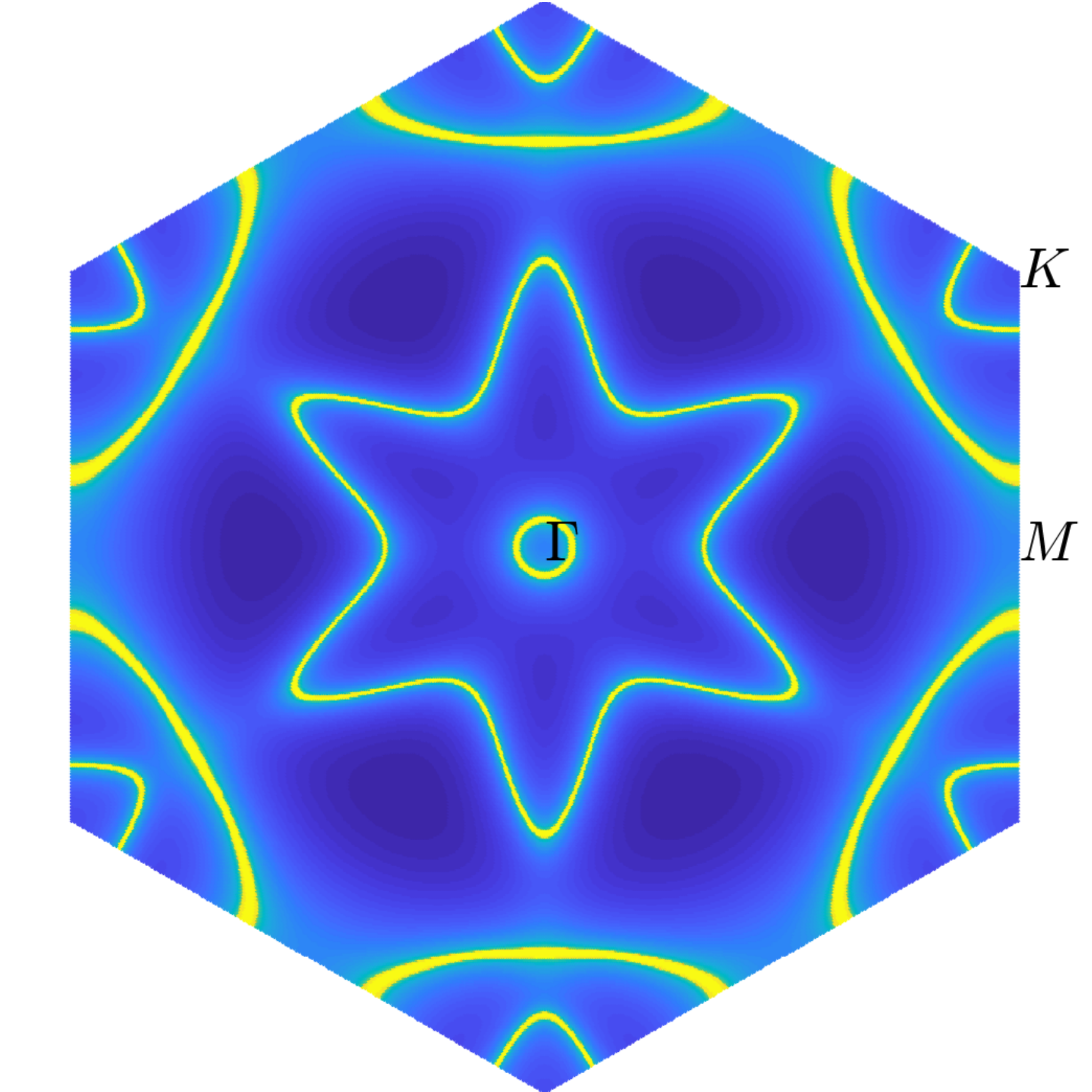}
    \caption{Left: Band structure of PdTe$_2$. The horizontal dashed line indicates the Fermi level. Right: Fermi surface contour in PdTe$_2$ for $k_z=0$ ($\Gamma$-$M$-$K$) plane.}
    \label{fig:dft_main_text}
\end{figure}
In order to specify the pairing states, let us first neglect the pockets around K and K$'$, which will later be taken into account when discussing the impact of disorder whenever relevant. With two Fermi surfaces, we have to distinguish between superconducting states that have the same sign (denoted by the superscript ``$++$'' in the following) and have opposite sign (superscript ``$+-$'') on the two Fermi surfaces. We, thus, have four different states, $A_{1g}^{++}$, $A_{1g}^{+-}$, $A_{1u}^{++}$, and $A_{1u}^{+-}$. The two singlet states $A_{1g}^{++}$ and $A_{1g}^{+-}$ have exactly the same transformation properties under all symmetries of the system, however, require different interactions driving the superconducting instability: the $A_{1g}^{++}$ state is expected to arise if the electron-phonon coupling (conventional pairing mechanism) or fluctuations of a time-reversal-symmetric collective electronic mode (such as charge-density fluctuations) provide the pairing glue \cite{ScheurerPRB2016}. Stabilizing the $A_{1g}^{+-}$ phase requires an effectively repulsive interaction between states on the two Fermi surfaces. As we will discuss below, also the behavior in the presence of disorder is different for these two singlet states. The two triplet states, $A_{1u}^{++}$ and $A_{1u}^{+-}$ also share the same symmetry properties and are expected to require repulsive intra-Fermi surface interactions. Which of the two is realized depends on the sign of the interactions between the two bands. 

\begin{figure*}
	\includegraphics[width=0.9\linewidth]{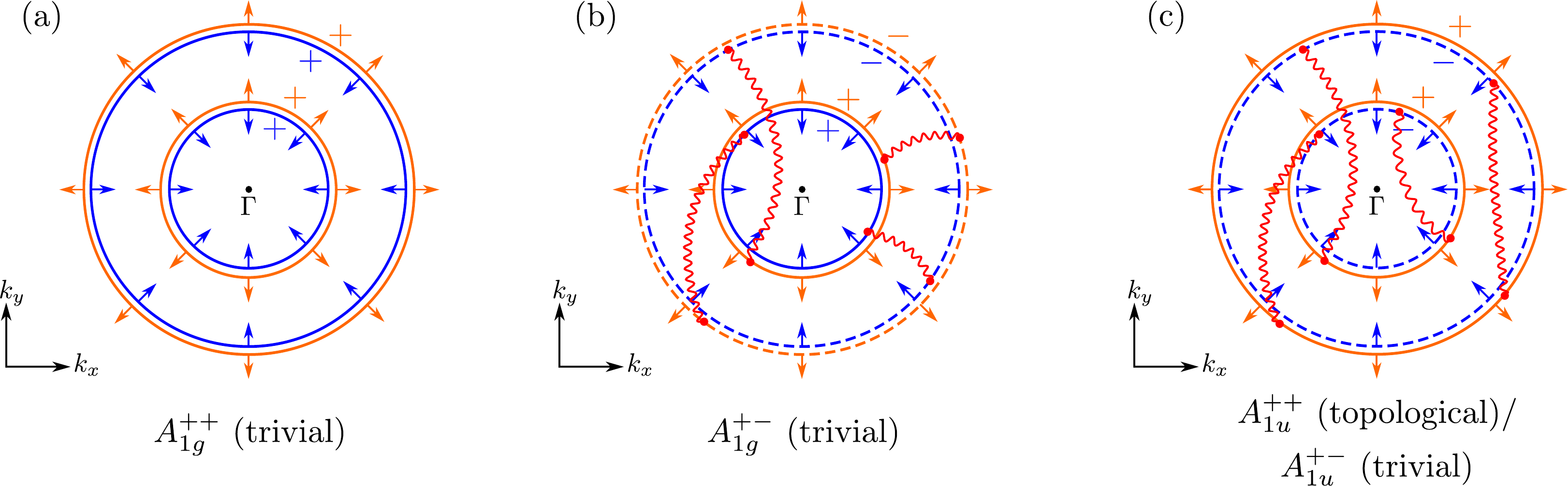}%
	\caption{(Color online) Illustration of the superconducting order parameter on the different Fermi surfaces around the $\Gamma$ point with solid (dashed) lines referring to positive (negative) sign of $\Delta_s(\vec{k})$ (see Eq.~\eqref{MatrixElementsOfSCOP}) and using the pseudospin-triplet basis defined in the main text, see~\equref{BasisChoice}. The arrows indicate the corresponding pseudospin polarizations taking the simplest form of the triplet vector, $\vec{d}_{\vec{k}}\propto (k_x,k_y,k_z)^T$, as an example. For clarity of the figure, we only show a two-dimensional cut in three-dimensional momentum space. While the system also has further Fermi surfaces around the K and K$'$ points, we here focus on those around the $\Gamma$ point as the extension to additional Fermi surfaces is straightforward and will be discussed later. We distinguish between candidate pairing states transforming under $A_{1g}$ with (a) equal ($A_{1g}^{++}$) and (b) opposite ($A_{1g}^{+-}$) sign of the singlet order parameter on the Fermi surfaces and (c) transforming under $A_{1u}$. While the singlet states are topologically trivial, the triplet state can be non-trivial, depending on the relation between the triplet vectors on the two Fermi surfaces. In the limit where the gap magnitude is isotropic and focusing on non-magnetic impurities, only scattering processes between the Fermi surfaces that are connected by red wiggly lines are pair breaking. The absence of wiggly lines in (a) indicates that the state $A_{1g}^{++}$ with constant gap is protected against all types of scattering processes, reproducing the Anderson theorem \cite{AndTh1,AndTh2,AndTh3}.} 
	\label{fig:ScatteringDifferentSCs} 
\end{figure*}

% Chiral basis:
\subsection{Chiral basis and topology}
Let us now determine the topological properties of the four pairing states, $A_{1g}^{++}$, $A_{1g}^{+-}$, $A_{1u}^{++}$, and $A_{1u}^{+-}$, and analyze which types of scattering processes are detrimental to superconductivity. For this, we will employ the following ``chiral'' basis: let $\vec{k}$ be in the vicinity of one of the doubly-degenerate Fermi surfaces of the system. We label the ones around the $\Gamma$ point as $n=1,2$ in the following [see Fig.~\ref{fig:dft_main_text}]. As a result of $P\Theta$ symmetry, the corresponding Bloch Hamiltonian will just be diagonal in pseudospin-space, $h_{\vec{k}} = \sigma_0 \epsilon_{\vec{k}n}$ with band energy $\epsilon_{\vec{k}n}$. Instead of choosing pseudospin up and down as basis states, we will take the non-degenerate eigenstates $\ket{\phi^s_{\vec{k}}}$, $s=\pm$, of the infinitesimally perturbed Hamiltonian
\begin{equation}
	h'_{\vec{k}} = \sigma_0 \epsilon_{\vec{k}n} + \alpha\,\vec{d}_{\vec{k}}\cdot \vec{\sigma}, \qquad \alpha \rightarrow 0^+, \label{BasisChoice}
\end{equation} 
satisfying $h'_{\vec{k}}\ket{\phi^s_{\vec{k}}} = \left(\epsilon_{\vec{k}n} + s\, 0^+\right)\ket{\phi^s_{\vec{k}}}$. Since the triplet vector obeys $\vec{d}_{\vec{k}} = -\vec{d}_{-\vec{k}} \in \mathbbm{R}^3 \setminus \{0\}$, the extra term in \equref{BasisChoice} lifts the degeneracy by infinitesimally breaking inversion while preserving time-reversal symmetry; it therefore holds, $\Theta \ket{\phi^s_{\vec{k}}} \propto \ket{\phi^s_{-\vec{k}}}$ and $P \ket{\phi^s_{\vec{k}}} \propto \ket{\phi^{-s}_{-\vec{k}}}$, where we suppressed momentum-dependent phase factors. Being infinitesimal, the second term in \equref{BasisChoice} has no physical consequences and should be viewed as a book-keeping tool that allows to define a convenient basis, $\{\ket{\phi^+_{\vec{k}}},\ket{\phi^-_{\vec{k}}} \}$. We will refer to this ``chiral'' basis as \textit{pseudospin-triplet basis}.

It is easy to see that the superconducting order parameters represented in this basis have the form ($s,s'=\pm$)
\begin{equation}
\widetilde{\Delta}_{ss'}(\vec{k})=\braket{\phi^{s'}_{\vec{k}}|\Delta_{\vec{k}}\left(i\sigma_2\right)^\dagger|\phi^s_{\vec{k}}} = \delta_{s,s'}\Delta_s(\vec{k}) \label{MatrixElementsOfSCOP}
\end{equation} 
with $\Delta_s(\vec{k}) = \Delta_0(\vec{k})$ for the singlet state transforming under $A_{1g}$ and $\Delta_s(\vec{k}) = s |\vec{d}_{\vec{k}}|$ for the triplet $A_{1u}$. Taking a finite value of $\alpha$ to make the different basis states $s=+$ (orange) and $s=-$ (blue) discernible, we show in \figref{fig:ScatteringDifferentSCs} a two-dimensional schematic cut of the Fermi surfaces around the $\Gamma$ point; furthermore, the sign of $\Delta_s(\vec{k})$ is indicated (solid/dashed lines) for the different candidate pairing states. Note that $\Delta_s(\vec{k}) \in \mathbbm{R}$ (without loss of generality) due to time-reversal symmetry and that $\Delta_s(\vec{k})$ has a fixed sign on each Fermi surface since we focus on fully gapped superconducting states without accidental nodes.

On top of providing a convenient way of illustrating the pairing states, the pseudospin-triplet basis also allows to easily infer their topological features: since all candidate states preserve $\Theta$, we have to view the superconductors as members of class DIII. Due to the infinitesimal perturbation in \equref{BasisChoice}, the Fermi surfaces are singly degenerate and the expression for the corresponding $\mathbbm{Z}$-valued topological invariant $\nu$ derived in \refcite{Invariants} can be applied. Using the result \cite{ScheurerPRB2016} that the Chern numbers of the two (infinitesimally split) Fermi surfaces of the Hamiltonian in \equref{BasisChoice} must be opposite, one can write
\begin{equation}
    \nu = \sum_n C^+_{n} \frac{\text{sign}(\Delta_+(\vec{k}_n))-\text{sign}(\Delta_-(\vec{k}_n))}{2}. \label{ExpressionForInvariant}
\end{equation}
In \equref{ExpressionForInvariant}, the summation over $n$ involves all pairs of infinitesimally split Fermi surfaces (in \figref{fig:ScatteringDifferentSCs}, there are two such pairs, $n=1,2$, of Fermi surfaces enclosing the $\Gamma$ point, i.e., pairs of orange and blue circles), $C^+_n$ is the Fermi surface Chern number of the $s=+$ member of the pair $n$, and $\vec{k}_n$ is an arbitrary momentum point on Fermi surface $n$.  

Recalling that we have $\Delta_+(\vec{k}) = \Delta_-(\vec{k})$ for the singlet states, we immediately find $\nu=0$ from \equref{ExpressionForInvariant}. Consequently, both $A_{1g}^{++}$ and $A_{1g}^{+-}$ are topologically trivial. This is different for the triplet states for which we have $\Delta_+(\vec{k}) = -\Delta_-(\vec{k}) \in \mathbbm{R}^+$ and, thus, $\nu = \sum_n C_n^+$. As each Fermi surface encloses an odd number of time-reversal invariant momenta (only the $\Gamma$ point), it follows that $C_{n}^+$ has to be odd \cite{Invariants} and, as such, non-zero. To illustrate this with a specific example, assume that $\vec{d}_{\vec{k}} \sim f(|\vec{k}|)(k_x,k_y,k_z)^T$ with some real-valued function $f$ describing the radial dependence of the triplet vector. If $f(|\vec{k}|)$ has the same sign on both Fermi surfaces, we find $C_{1}^+=C_{2}^+=1$ and the topologically nontrivial value $\nu=2$; this is the $A_{1u}^{++}$ state (which we define more generally by $C_{1}^+\neq -C_{2}^+$). In the case of the $A_{1u}^{+-}$ state, $f(|\vec{k}|)$ changes sign leading to $C_{1}^+=-C_{2}^+=1$ and $\nu=0$ (topologically trivial). 

To summarize, the singlets $A_{1g}^{++}$, $A_{1g}^{+-}$ are topologically trivial. The triplet states can be topologically non-trivial with non-zero, even value of the invariant $\nu$ in \equref{ExpressionForInvariant} depending on the form of the $\vec{d}_{\vec{k}}$ vector: for instance, if $\vec{d}_{\vec{k}} \sim f(|\vec{k}|)(k_x,k_y,k_z)^T$, the state will be topological (trivial) if $f$ does not change sign, corresponding to $A_{1u}^{++}$ (changes sign, corresponding to $A_{1u}^{+-}$), between the two Fermi surfaces around the $\Gamma$ point.

\subsection{Generalized Anderson theorem}
\label{sec:generalized_Anderson_theorem}
To discuss the expected impact of impurity scattering for the different candidate pairing states, we use the \textit{generalized Anderson theorem} of \refcite{Scheurer2016}: let $\hat{W}$ be the matrix comprising the impurity-induced scattering amplitudes $(\hat{W})_{\vec{k}\alpha,\vec{k}'\alpha'} = \braket{\vec{k}\alpha|W|\vec{k}'\alpha'}$ between the single-particle states $\ket{\vec{k}\alpha}$ and $\ket{\vec{k}'\alpha'}$ of the clean normal state Hamiltonian $\left(h_{\vec{k}}\right)_{\alpha\alpha'}$ for which crystal momentum $\vec{k}$ is still a good quantum number. Here, $\alpha$ and $\alpha'$ label all remaining relevant degrees of freedom, such as spin and various orbitals. Similarly, we introduce the matrix elements $\hat{D}_{\vec{k}\alpha,\vec{k}'\alpha'} = \delta_{\vec{k},\vec{k}'}\braket{\vec{k}\alpha|\Delta_{\vec{k}}T^\dagger|\vec{k}'\alpha'}$ of the pairing order parameter $\Delta_{\vec{k}}$, where $T$ is the unitary part of the time-reversal operator, $\Theta = T \mathcal{K}$ ($\mathcal{K}$ denotes complex conjugation); for instance, we have $T=i\sigma_2$ in the (pseudo)spin basis described above. As long as the electronic states are still delocalized in the vicinity of the Fermi  ($k_F \ell \gg 1$), the superconducting critical temperature is unaffected by the presence of disorder if both
\begin{subequations}\begin{equation}
    \left[\hat{h},\hat{D}\right]_- = 0, \label{GATCond1}
\end{equation}
where $\hat{h}_{\vec{k}\alpha,\vec{k}'\alpha'} = \delta_{\vec{k},\vec{k}'} \left(h_{\vec{k}}\right)_{\alpha\alpha'}$, and
\begin{equation}
    \left[\hat{W} , \hat{D} \right]_{-t_W} = 0  \label{GATCond2}
\end{equation}\label{GeneralizedATCond}\end{subequations}
hold. Here $[\hat{A},\hat{B}]_\pm = \hat{A}\hat{B} \pm \hat{B}\hat{A}$ and $t_W = +$ ($t_W=-$) for non-magnetic (magnetic) disorder, i.e., $\Theta W \Theta^\dagger = t_w W$. While this general form of Anderson's theorem has been derived in \refcite{Scheurer2016}, we present a compact justification in  \appref{GeneralizedAT} for convenience and to provide a more formal definition of the notation used. We emphasize that this result holds for arbitrary momentum dependence of the order parameter. One advantage of the condition (\ref{GeneralizedATCond}) for the validity of the generalized Anderson theorem is that it only depends on the (anti)commutation relations of the order parameter, the normal state Hamiltonian, and the disorder potential and, hence, can be readily tested in any single-particle basis. 

The pseudospin triplet basis introduced above is most convenient for us. In this low-energy description, only the matrix elements of the superconducting order parameter within each band (associated with the Fermi surfaces $n=1,2$ or $n=1,2,3,4$ depending on whether we neglect or take into account the Fermi surfaces around the K, K$'$ points) are kept and, hence, \equref{GATCond1} is automatically satisfied. 
Using the form in \equref{MatrixElementsOfSCOP} of the matrix elements of the superconducting order parameter in the pseudospin-triplet basis, the remaining second condition (\ref{GATCond2}) simply becomes 
\begin{subequations}
\begin{equation}
    C_{\vec{k}s,\vec{k}'s'} = 0,  \qquad \forall\, \vec{k},\vec{k}'\hspace{-0.1em},s,s', \label{CommutatorInPSBasis}
\end{equation}
where we introduced (no summation over repeated indices)
\begin{equation}
    C_{\vec{k}s,\vec{k}'s'}:=\left(\Delta_{s}(\vec{k}) - t_W \Delta_{s'}(\vec{k}') \right) \braket{\phi_{\vec{k}}^s|W|\phi_{\vec{k}'}^{s'}}.\label{DefinitionOfCorrelatorInBasis} 
\end{equation}\label{ATInChiralBasis}\end{subequations}
This means that the superconducting state is protected against non-magnetic (magnetic) disorder if scattering matrix elements are non-zero only between single-particle states $\ket{\phi_{\vec{k}}^s}$ and $\ket{\phi_{\vec{k}'}^{s'}}$ for which the superconducting order parameter $\Delta_s(\vec{k})$ has the same value (same magnitude but opposite sign). Therefore, we can readily read off which scattering events are pair breaking for the different superconducting states in \figref{fig:ScatteringDifferentSCs}. 

The impurities induced by electron irradiation are non-magnetic, and we thus focus on $t_W = +$ in the following. Let us first discuss the case of a single, isotropic gap energy scale, i.e., that $\Delta_s(\vec{k})$ only depends on $s$ and the Fermi surface where $\vec{k}$ is located at and that the magnitude $|\Delta_s(\vec{k})|$ is constant; we will examine the general case below. 
The order parameter is then identical on all Fermi surfaces for the $A_{1g}^{++}$ state. As we read off from \equref{ATInChiralBasis}, its critical temperature will be unaffected by any time-reversal-symmetric (TRS) impurity as long as the mean-free path is much larger than the Fermi wavelength $k_F \ell \gg 1$ (no localization). This reproduces the well-known conventional Anderson theorem for a multiband system~\cite{AndTh1,AndTh2,AndTh3}. 
This is different for the $A_{1g}^{+-}$ state, which is prone to scattering between the small and large pockets as a consequence of the sign change of $\Delta_s(\vec{k})$ between the two Fermi surfaces (see \figref{fig:ScatteringDifferentSCs}(b)). Similarly, both triplet states $A_{1u}^{++}$ and $A_{1u}^{+-}$ are suppressed by scattering events between four out of the ten pairs of Fermi surfaces (see ~\figref{fig:ScatteringDifferentSCs}(c)).

\subsection{Rate of $T_c$ suppression}
\label{sec:rate_Tc_suppression}
Let us now show that the commutator $C_{\vec{k}s,\vec{k}'s'}$ introduced in~\equref{DefinitionOfCorrelatorInBasis} determines the rate of suppression of $T_c$ with increasing scattering rate. As shown in detail in Appendix~\ref{WeakDisorderLimit}, the change $\delta T_c := T_c - T_{c,0}$ of the critical temperature $T_c$ relative to its clean value, $T_{c,0}$, can be expressed entirely in terms of $C_{\vec{k}s,\vec{k}'s'}$ as
\begin{subequations}
\begin{equation}
    \delta T_{c}/T_{c,0} \sim -\frac{\pi}{4T_{c,0}}\tau^{-1} \, \zeta \, ,
    \label{eq:Tc_suppression}
\end{equation}
in the asymptotic limit of low disorder configurations (small scattering rate $\tau^{-1}\rightarrow 0$). All non-universal features that depend, for instance, on details of the superconducting order parameter and impurity potential enter the dimensionless ``sensitivity parameter'' 
\begin{equation}
    \zeta =  \frac{\sum_{\vec{k},\vec{k}'}^{\text{FS}}\sum_{s,s'} |C_{\vec{k}s,\vec{k}'s'}|^2}{2\, \text{tr}[W^\dagger W]\sum_{\vec{k}}^{\text{FS}}\sum_s |\Delta_s(\vec{k})|^2} \,.
    \label{ZetaExpression}
\end{equation}\label{SuppresionOfTcMainText}\end{subequations}
The parameter $\zeta$ involves the Fermi surface average of the commutator in \equref{DefinitionOfCorrelatorInBasis}. Here, we have written $\sum_{\vec{k}}^{\text{FS}} \dots  \equiv \frac{1}{N_\Lambda} \sum_n\sum_{\vec{k},|\epsilon_{\vec{k}n}|<\Lambda} \dots $, with momentum cutoff $\Lambda$ and total number of momentum points involved $N_{\Lambda}=\sum_{\vec{k}}^{\text{FS}}$. The trace in the denominator is over all internal degrees of freedom (spin, orbitals) of the impurity potential to ensure proper normalization of $\zeta$. Within our conventions, it holds $\zeta = 1$ in the AG case, i.e., for magnetic impurities in a single-band, isotropic spin-singlet superconductor. We note that a special case (momentum-independent superconducting order parameter) of the observation that non-universal features can be absorbed into an effective scattering involving a commutator has been recently derived \cite{arXivGAT}; however, there are a few subtleties with the application of this commutator to be discussed below.

As follows from comparing the slopes in \figref{Tcgamma} of the AG law (black dashed line) and that of the curve  measured for PdTe$_2$ (red solid line), we experimentally determine $\zeta \approx 1/16$ for PdTe$_2$. This means that the suppression of $T_c$ with disorder is weaker by approximately a factor of $16$ compared to magnetic impurities in a superconductor with a momentum independent order parameter on the Fermi surface. While $\zeta \simeq 1/16$ is small and one may naively conclude that the order must be $A_{1g}^{++}$, it is important to perform a quantitative analysis, taking the Fermi surface geometry properly into account. In the following, we therefore analyze in detail under which conditions the different fully gapped pairing options are consistent with the measured $T_c$ suppression rate in PdTe$_2$.

\subsection{Singlet pairing}
\label{subsec:singlet_pairing}
Let us begin with singlet pairing and first note that we estimate $k_F \ell $ to be of order $10^3$ based on our resistivity measurements (see \appref{subsec:electrical_resistivity}). Due to the sizeable value of $k_F \ell$, we expect localization effects to play only a minor role in the suppression of $T_c$ in the three-dimensional system PdTe$_2$. In other words, a suppression of $T_c$ with disorder can only be realized if the conditions of the generalized Anderson theorem in \equref{ATInChiralBasis} are violated. The finite change of $T_c$ with disorder we observe is only consistent with either the $A_{1g}^{+-}$ state or the $A_{1g}^{++}$ superconductor with momentum dependent gap function $|\Delta_s(\vec{k})|$. In fact, we generically expect a momentum dependent gap, however, the observed temperature dependence of the London penetration depth is only consistent with moderately weak anisotropies of the gap function.  

In order to evaluate the parameter $\zeta$ using \equref{ZetaExpression} for a specific pairing state such as singlet pairing, $\Delta_s(\vec{k}) = \Delta_0(\vec{k})$ with arbitrary momentum dependence, we need to make an assumption about the scattering matrix $\hat{W}$. Here, we consider the simplest case of $W=\sigma_0$, which corresponds to point-like scalar disorder without any momentum dependence in the pseudospin basis. Note that this assumption on the impurity potential does not take into account the multi-orbital nature of the system that might further suppress the impact of impurities on $T_c$~\cite{DisorderSOCFu,OurDisorderSOC,BrydonScattering}, as we will discuss in~\secref{TripletPairing}. Therefore, the following estimates on the momentum dependence and the degree of anisotropy of the order parameter $\Delta_0(\vec{k})$, which are based on the requirement to yield $\zeta \simeq 1/16$, should be viewed as lower bounds. From \equref{ZetaExpression}, we find
\begin{equation}
    \zeta = \frac{\braket{|\Delta_0|^2}_{\text{FS}}-|\braket{\Delta_0}_{\text{FS}}|^2}{2\braket{|\Delta_0|^2}_{\text{FS}}}, \label{GapAnisotropy}
\end{equation}
where $\braket{\cdots }_{\text{FS}} \equiv \sum_{\vec{k}}^{\text{FS}} \dots $ denotes averaging over the Fermi surface. In accordance with previous results ~\cite{GolubovMarzin,KoganScattering}, the suppression of $T_c$ can be expressed in terms of the variance of the gap on the Fermi surface. 

\begin{figure}
   \centering
    \includegraphics[width=\linewidth]{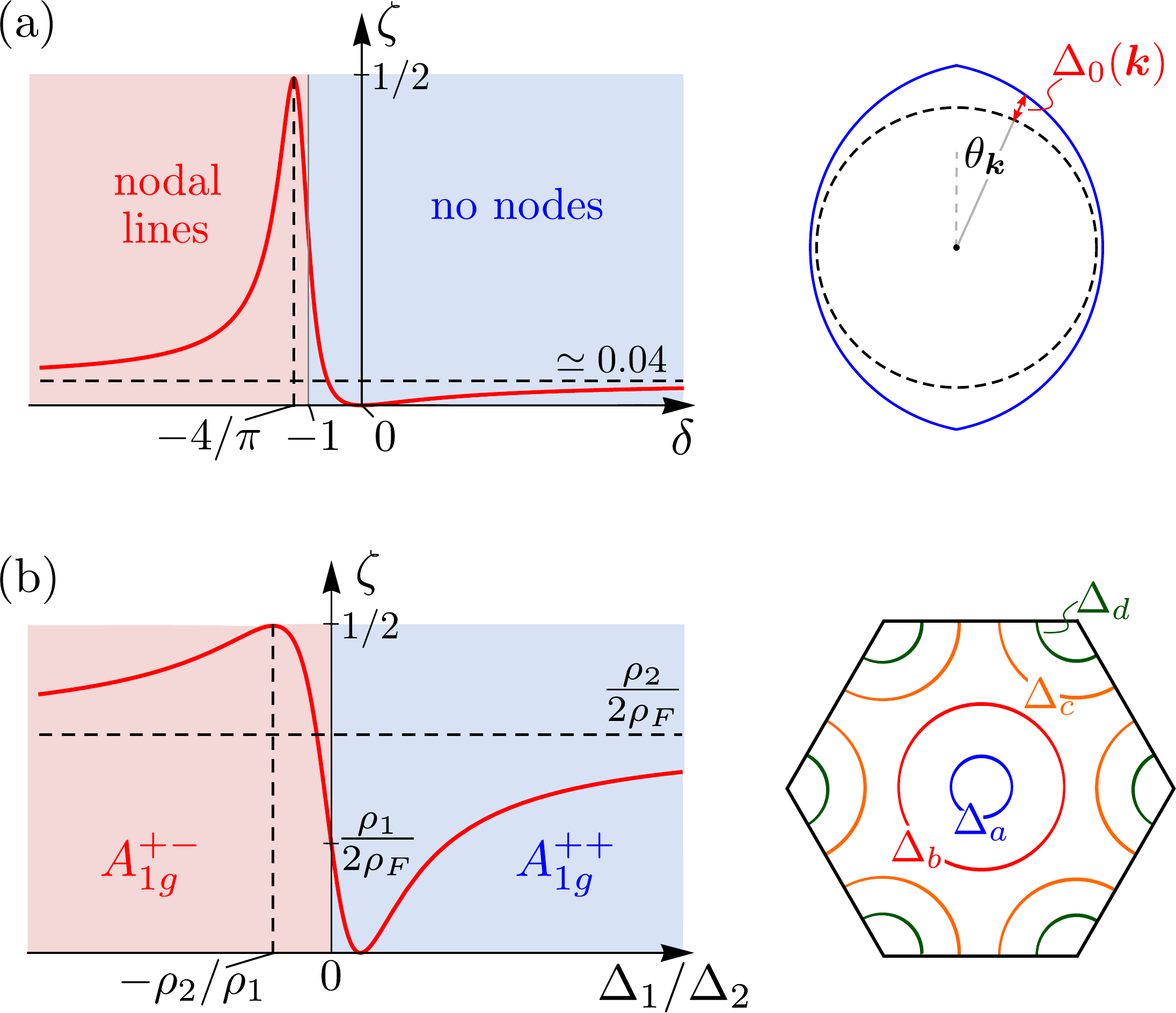}
    \caption{Sensitivity $\zeta$ in \equref{SuppresionOfTcMainText} of singlet states with momentum dependent order parameters to non-magnetic disorder. In (a), we show $\zeta$ (left panel) for a singlet state with order parameter $\Delta_s(\vec{k}) = \Delta_0(\vec{k})$ varying on a single, spherical Fermi surface as parametrized in \equref{DefinitionOfGapVariation} and illustrated in the right panel. The sensitivity $\zeta$ of a state with constant gap on each Fermi surfaces but two different values, $\Delta_n$, of $\Delta_0(\vec{k})$ is shown in (b), left panel, as a function $\Delta_1/\Delta_2$ together with its dependence on the individual, $\rho_n$, and total, $\rho_F = \sum_n \rho_n$, density of states. Two relevant applications to the Fermi surfaces of PdTe$_2$, shown schematically in the right panel, are  $\Delta_a=\Delta_b=\Delta_c = \Delta_1$, $\Delta_d=\Delta_2$, and $\Delta_a=\Delta_b=\Delta_1$, $\Delta_c=\Delta_2$. The former gives rise to an isotropic $A_{1g}^{+-}$ state, the latter to an $A_{1g}^{++}$ state with $\Delta_1/\Delta_2 = 2.1$ as discussed in the main text.}
    \label{fig:A1gDetrimentalProcesses}
\end{figure}

Two possible types of anisotropy give rise to non-zero $\zeta$ in \equref{GapAnisotropy}: \textit{(i)} a variation of the order parameter magnitude on the Fermi surfaces, and \textit{(ii)} the order parameter is constant on each Fermi surface but has a different size on distinct Fermi surfaces. While both anisotropy types may be present simultaneously in the experimental system, we discuss them separately for clarity. 

\subsubsection{Anisotropic gap around Fermi surface}
In case \textit{(i)}, it is sufficient to focus on one Fermi surface around the $\Gamma$ point that we assume to be spherical; its momenta satisfy $|\vec{k}| = k_F$. Let us for simplicity only take the leading order, i.e., lowest harmonic, momentum dependence of the order parameter into account:
\begin{equation}
	\Delta_0(\vec{k}) = \Delta_0\left(1 + \delta \sin (\theta_{\vec{k}}) \right), \label{DefinitionOfGapVariation}
\end{equation}
where we have used spherical coordinates, $\vec{k} = k_F\, (\sin \theta_{\vec{k}} \cos \phi_{\vec{k}},\sin \theta_{\vec{k}} \sin \phi_{\vec{k}},\cos \theta_{\vec{k}})^T$ with the three-fold rotation symmetry of the point group $D_{3d}$ along the $k_z$ axis. A single, dimensionless parameter $\delta \in \mathbbm{R}$ parametrizes the gap variation, see right panel in \figref{fig:A1gDetrimentalProcesses}(a). Note that nodes are present if and only if $\delta \leq -1$. From \equref{GapAnisotropy}, we obtain after straightforward algebra and converting all sums into integrals that the sensitivity parameter is given by
\begin{equation}
\zeta = \frac{(32-3\pi^2)\delta^2}{16(6+3\pi \delta + 4 \delta^2)}\,. \label{ZetaInTermsOfdelta}
\end{equation}
This result is visualized in the left panel of~\figref{fig:A1gDetrimentalProcesses}(a). We first note that the maximal value of $\zeta$ that can be realized with positive $\delta$ is quite small and given by $1/2-3\pi^2/64\approx 0.037$. As follows from \equref{DefinitionOfGapVariation}, large $\delta >0$ means near nodes at the ``poles'' of the Fermi surface  ($\theta_{\vec{k}}=0$). Larger values of $\zeta$ occur for negative $\delta$, which corresponds to having nodal lines ($\delta<-1$) or near nodal lines  ($-1<\delta<0$) in the vicinity of the ``equator'' of the Fermi surface ($\theta_{\vec{k}}=\pi/2$). For our experimentally observed value of $\zeta \approx 1/16$ in PdTe$_2$, there are two possible values $\delta \approx -0.73$ or $\delta \approx -5.02$ according to \equref{ZetaInTermsOfdelta}. While the latter corresponds to accidental nodal lines, the former has a non-zero gap on the entire Fermi surface but a significant degree of anisotropy with $
\min_{\vec{k}} |\Delta_0(\vec{k})|/\max_{\vec{k}} |\Delta_0(\vec{k})| \approx 0.27$. 
Both of these values of $\delta$ are not consistent with the measured temperature dependence of the penetration depth in Fig.~\ref{rhos}, which can be well described by a single gap energy scale and is known to exhibit a different temperature dependence in the presence of such significant gap anisotropy~\cite{ProzorovKogan2011}.  The variation of the magnitude of the order parameter on the Fermi surfaces seems unlikely to be the dominant source of the observed non-zero value of $\zeta$.

\subsubsection{Different constant gaps on multiple Fermi surfaces} 
Let us now focus on scenario \textit{(ii)} where $\Delta_0(\vec{k})$ is constant on each Fermi surface, $n=1,2,\dots N$, but can take on different magnitudes on the different Fermi surfaces, i.e., $\Delta_0(\vec{k}_n) = \Delta_n$. Denoting the total density of states of Fermi surface $n$ by $\rho_n$, we immediately obtain from \equref{GapAnisotropy}
\begin{equation}
\zeta = \frac{1}{2} \left[1-\frac{\left(\sum_n \rho_n \Delta_n \right)^2}{\left(\sum_n \rho_n \Delta^2_n\right) \sum_n \rho_n}\right] \label{SecondExpressionForZeta}
\end{equation}
The behavior of $\zeta$ for the case of two Fermi sheets, $N=2$, is shown in \figref{fig:A1gDetrimentalProcesses}(b), left panel. Quantitative predictions require knowledge of the density of states $\rho_n$ at the different Fermi surfaces.  Using DFT calculations, we find for PdTe$_2$ that $\rho_a = 0.010\,\text{eV}^{-1}$, $\rho_b = 0.39\,\text{eV}^{-1}$, $\rho_c = 0.91\,\text{eV}^{-1}$, and $\rho_d = 0.045\,\text{eV}^{-1}$ (with labeling of Fermi surfaces shown in  Fig.~\ref{fig:A1gDetrimentalProcesses}(b), right panel).

Let us for simplicity first concentrate on the case of only two different gap values $\Delta_1$ and $\Delta_2$. We consider the most general case with possibly four different gaps in Appendix~\ref{app_sec:gap_anisotropies}, and show that this does not change our conclusions. For two gap sizes $\Delta_1, \Delta_2$ there are in total eight different possibilities of how these can be distributed over the four Fermi sheets $a,b,c,d$. The resulting gap ratios $\Delta_1/\Delta_2$ that are consistent with $T_c$ suppression slope of $\zeta = 1/16$ are listed in Appendix~\ref{app_sec:gap_anisotropies}. The most isotropic state we find is a $A_{1g}^{+-}$ state with a sign change between the small electron pocket $d$ and the other three pockets
\begin{equation}
\Delta_a = \Delta_b = \Delta_c = \Delta_2, \Delta_d = \Delta_1
\end{equation}
Such an isotropic state is a plausible option, since it is consistent with the observed isotropic temperature dependence of the penetration depth $\lambda$.

Interestingly, we find that the $A_{1g}^{++}$ states always exhibit a larger degree of anisotropy. The smallest anisotropy we find is about $\Delta_1/\Delta_2 \approx 2.1$ ($2.0$ when we allow for four different gap sizes). As shown in Appendix~\ref{app_sec:gap_anisotropies} 
this occurs for a number of combinations of how the two gap sizes $\Delta_{1,2}$ are distributed over the four Fermi sheets. Importantly, such a large degree of  anisotropy is expected to be visible in the temperature dependence of the London penetration depth that we find to be well described by a single gap energy scale (see Sec.~\ref{sec:results})~\cite{ProzorovKogan2011}. We conclude that the $A_{1g}^{++}$ solutions are inconsistent with our observations that $\lambda(T)$ at least under the (natural) assumption that intra- and inter-band scattering is equally strong. We note that the observed slow (and seemingly saturating) $T_c$ suppression at larger values of $\gamma^\lambda$ (see Fig.~\ref{Tcgamma}) on the other hand indicates extremely robust superconducting pairing. We suggest to further investigate this question in future work. 

To summarize, the most likely scenario in the $A_{1g}$ channel is an unconventional $A_{1g}^{+-}$ state with sign change between the small electron pocket $d$ around the $K, K'$ points and the other Fermi sheets $\{a, b, c\}$. This state is completely isotropic with a ratio of gap magnitudes given by  $|\Delta_1|/|\Delta_2| \approx 1$. Any (conventional) $A_{1g}^{++}$ pairing state has a significant degree of gap anisotropy. The minimal gap ratio we find is $\Delta_1/\Delta_2 \approx 2$, making these states inconsistent with our observations of a London penetration depth $\lambda(T)$ that is well described by a single gap energy scale.  

\subsection{Triplet pairing}
\label{TripletPairing}
In this subsection, we consider $T_c$ suppression with disorder for the $A_{1u}$ triplet state. From a direct comparison of the curve for PdTe$_2$ and that of the naive AG law in \figref{Tcgamma}, one is tempted to conclude that the observed suppression of $T_c$ in PdTe$_2$ is too weak to be consistent with a triplet pairing state. Indeed, as readily follows from \equref{SuppresionOfTcMainText}, we obtain $\zeta = 1/2$ for any triplet state in the pseudospin approximation for the case of a single scattering rate arising from the most detrimental assumption of an impurity potential that is diagonal and momentum-independent in the pseudospin basis, $\braket{\phi_{\vec{k}}^s|W|\phi_{\vec{k}'}^{s'}} = W_0 \delta_{ss'}$. This applies to both the $A_{1u}$ and also the third candidate pairing state $E_u(1,0)$~\cite{PRB} and agrees with previous results, e.g., \refscite{Maekawa,Puchkaryov}. Intuitively, $\zeta = 1/2$ results from the fact that only scattering processes \emph{between} the infinitesimally split Fermi surfaces of $h_{\vec{k}}$ in \equref{BasisChoice} are pair breaking, while also those \emph{within} each of the Fermi surfaces are detrimental to superconductivity in the AG case (magnetic impurities and momentum independent singlet superconductor), see \equref{DefinitionOfCorrelatorInBasis}. To reconcile the observed suppression of $T_c$ in PdTe$_2$ in \figref{Tcgamma} with a triplet pairing state, we therefore, need approximately an additional factor of $8$ of reduction of $\zeta$.

As already alluded to above, such a reduction could in principle result from a suppression of the scattering matrix elements $\braket{\phi_{\vec{k}}^s|W|\phi_{\vec{k}'}^{s'}}$ that is related to the fact that we are working in the pseudospin basis. Consequently, even for the simplest case of $\vec{d}_{\vec{k}} \sim (k_x,k_y,k_z)$, the arrows in \figref{fig:ScatteringDifferentSCs} refer to pseudospin polarizations rather than just spin polarizations. Due to the significant spin-orbit coupling, these will generally be momentum-dependent admixtures of spin and the Pd $4d$ and Te $5p$ orbitals relevant in the vicinity of the Fermi energy. This is confirmed within our detailed DFT calculations, which are presented in Appendix~\ref{sec_app:DFT}.
Such spin-orbital mixing is expected to suppress scattering which can lead to a significant parametric enhancement \cite{DisorderSOCFu,OurDisorderSOC,BrydonScattering} of the critical scattering strength relative to the naive AG law \cite{AG}. 
Intuitively, the reason for this suppression is that the angular dependence of the orbital- and spin-polarization of the Bloch states can lead to strongly suppressed matrix elements of the impurity potential, as has been discussed in detail in \refcite{OurDisorderSOC}. 
We note, however, that the required reduction in PdTe$_2$ is by a factor of eight, which is rather large. This makes the triplet state a rather unlikely candidate pairing state. 

Nevertheless, it is worth emphasizing that under certain conditions a fully gapped spin-triplet SC can enjoy an Anderson theorem; for instance, this is possible if the order parameter is momentum independent while its non-trivial symmetry properties are accounted for by its orbital structure. This was demonstrated by Michaeli and Fu in \refcite{DisorderSOCFu} by an explicit model calculation. This phenomenon can be readily understood from the generalized Anderson theorem (\ref{GeneralizedATCond}), applied in the orbital basis: assuming that the impurity potential is trivial in orbital space, we have $\hat{W}_{\vec{k}\alpha,\vec{k}'\alpha'} = \delta_{\alpha,\alpha'} f_{\vec{k},\vec{k}'}$, which always commutes with a momentum independent pairing potential, $\hat{D}_{\vec{k}\alpha,\vec{k}'\alpha'} = \delta_{\vec{k},\vec{k}'} D_{\alpha\alpha'}$. While \equref{GATCond2} is automatically satisfied, we have to keep in mind that also \equref{GATCond1} has to hold~\cite{Note1}. %\footnote{One might think that the generalized Anderson theorem should hold upon projecting the order parameter onto the low-energy basis, without the need of an extra symmetry. Then, indeed, \equref{GATCond1} will be satisfied. However, the projection will affect the commutation properties between the disorder potential and the order parameter; without additional symmetries of the Hamiltonian, \equref{GATCond2} will become invalid as can be easily verified.}. 
Therefore, the generalized Anderson theorem only applies if the normal state Hamiltonian has the additional symmetry, $[h_{\vec{k}},D] = 0$. This is the reason why \refcite{DisorderSOCFu} only finds a constant $T_c$ in the presence of a chiral symmetry and agrees with the very recent result \cite{BrydonScattering}. 

We note in passing that a special case (the limit without momentum dependence of the order parameter) of the second condition (\ref{GATCond2}) has recently been re-derived in \refcite{arXivGAT}. In that work, however, the first condition (\ref{GATCond1}) was not properly taken into account. We believe that this is the reason why the results of \refcite{arXivGAT} disagree with ours and those of \refscite{DisorderSOCFu,BrydonScattering}. In particular, the crucial first condition (\ref{GATCond1}) can generally be shown to be incompatible with the presence of nodes that are induced by the projection of a momentum independent pairing potential on the Fermi surface. This contradicts the claim of a generalized Anderson theorem for nodal superconductors made in \refcite{arXivGAT}. 

In contrast, the candidate $A_{1u}$ triplet states in PdTe$_2$ possess a full gap and can thus, in principle, fulfill the first condition (\ref{GATCond1}) of the generalized Anderson theorem. Our observation that the rate of suppression of $T_c$ with disorder is smaller than what we would expect from the AG law does therefore not necessarily rule out the triplet pairing state $A_{1u}^{++}$, or $A_{1u}^{+-}$, yet requires some fine-tuning of the microscopic orbital structure of the order parameter and the normal state Hamiltonian. Depending on these microscopic details, spin-orbit coupling can be responsible for the additional suppression of the sensitivity parameter $\zeta$ by a factor of $8$. Interestingly, our DFT results presented in Appendix~\ref{sec_app:DFT} demonstrate that all four bands exhibit a substantial amount of mixing between states of opposite parity at the Fermi surface. This is a necessary (though not sufficient) condition for the protection mechanism outlined by Michaeli and Fu~\cite{DisorderSOCFu}. Further work, which takes microscopic details of the orbital content of the bands, possible pairing interactions and the disorder potential into account, is necessary to investigate this intriguing possibility. Here we conclude that to explain the observed slow $T_c$ suppression requires making more assumptions about microscopic details in case of the $A_{1u}$ triplet states than for the $A_{1g}$ singlet states. The triplet states are therefore less likely realized in PdTe$_2$. 

\section{Conclusions}
\label{sec:conclusions}
We report on the impact of electron irradiation on the properties of the normal and superconducting state in PdTe$_2$. Our detailed study reveals that electron irradiation controllably tunes the non-magnetic scattering rate $\tau^{-1}$ without affecting carrier densities. The superconducting state remains fully gapped under irradiation, but its transition temperature $T_c$ is suppressed at a rate of about $\zeta \simeq 1/16$ compared to $\zeta = 1$ found for the Abrikosov-Gorkov law. We find the temperature dependence of the London penetration depth to be well described by a single gap energy scale, which is only consistent with a rather weak degree of anisotropy of the absolute value of the superconducting gap across the Fermi surfaces. 

We use this information to infer properties of the superconducting pairing states, and in particular show under which conditions the different possible candidate states yield a $T_c$ suppression rate consistent with experiment. One of our main conclusions is that the powerful probe of controlling the amount of disorder using electron irradiation must be combined with a thorough theoretical analysis in order to draw correct conclusions about the nature of the superconducting pairing state. Our analysis is based on the generalized Anderson theorem of \refcite{Scheurer2016} for multi-band superconductors, which can be expressed in algebraic form in terms of (anti)commutators of the gap, the disorder potential, and the normal state Hamiltonian. We use this powerful formulation to demonstrate that the suppression rate is governed by the Fermi surface average of the exact same commutator. Our general formulation agrees with previous results derived in various limits. A weak violation of the generalized Anderson theorem conditions therefore leads to a slow suppression rate. This situation can both apply to anisotropic spin-singlet superconductors as well as to (fully gapped) triplet paired states in the presence of strong spin-orbit coupling. Importantly, our concise formulation makes it easily applicable to other situations. 

It is important to note that quantitative predictions of our theory depend on microscopic details such as the disorder potential matrix elements between different momentum and orbitals. Here, we focus on the (most agnostic) assumption of a momentum-independent disorder potential that acts diagonal in the pseudo-spin basis. For the $A_{1g}^{+-}$ case, this corresponds to a ratio of inter- to intra-band scattering rates equal to one. The results also depend on the density of states at the different Fermi pockets, which we obtain using density-functional theory. 

Under these assumptions, we conclude that the most likely pairing state is the unconventional $A_{1g}^{+-}$ state with a different sign of the gap on the inner hole pocket and the three other Fermi pockets. We find that such the $A_{1g}^{+-}$ state can be completely isotropic $|\Delta_1|/|\Delta_2 \approx 1.0$ and exhibit a $T_c$ suppression with slope $\zeta = 1/16$ as we experimentally observe. In contrast, a conventional $A_{1g}^{++}$ state must have a gap anisotropy of at least $\Delta_1/\Delta_2 \approx 2$. This is not consistent with our results that $\lambda(T)$ can we well described by a single gap energy scale. We thus conclude that an anisotropic $A_{1g}^{++}$ state is only consistent with the experimental data if the ratio of intra- to inter-band scattering is much smaller than one. Further work should try to reveal details of the scattering defect potential introduced by electron irradiation, in particular its orbital structure.

\section{Acknowledgements}
We thank Vladimir G. Kogan for illuminating discussion of generalized pair-breaking theories. We thank Morgan Masters, Joshua Slagle and Victor Barrena Escolar for support during the crystal growth. We also acknowledge discussions with D.~C.~Cavanagh and P.~M.~R.~Brydon. The experimental work was supported by the U.S. Department of Energy (DOE), Office of Basic Energy Sciences, Division of Materials Sciences and Engineering. The experimental research was performed at Ames Laboratory. Ames Laboratory is operated for the U.S. DOE by Iowa State University under Contract No.~DE-AC02-07CH11358. 
Development of thermal conductivity mobile thermal conductivity (MTC) setup was supported by the Laboratory Research and Development Program of The Ames Laboratory under the U.S. Department of Energy Contract No. DE-AC02-07CH11358.
N.H.J. was supported by the Gordon and Betty Moore Foundation's EPiQS Initiative (Grant No. GBMF4411).
Y.L. and L.K. acknowledge the support from the U.S. DOE Early Career Research Program.
M.S.S. acknowledges support from the German National Academy of Sciences Leopoldina through grant LPDS 2016-12 and from the National Science Foundation under Grant No.~DMR-1664842. P.P.O. acknowledges support from Iowa State University and Ames Laboratory Startup Funds.

\appendix
\section{Generalized Anderson theorem}\label{GeneralizedAT}
To be self-contained and to further clarify all quantities entering the generalized Anderson theorem used in the main text, we will provide a compact derivation of \equref{GeneralizedATCond} in this appendix. The following approach is adapted from
\refcite{Scheurer2016}, where also a diagrammatic proof can be found. We note that a special case of the following simplified argument has also been published in \refcite{Hoyer2015}. 

As opposed to the main text, we use second quantization with $c^\dagger_{\vec{k}\alpha}$ denoting the creation operator for an electron with momentum $\vec{k}$ and spin, orbital, etc.~quantum numbers labelled by the multi-index $\alpha$. We consider a general superconductor in $d$ spatial dimensions with mean-field Hamiltonian
\begin{align}\begin{split}
H_{0} &= \sum_{\vec{k}} c^\dagger_{\vec{k}\alpha} \left(h_{\vec{k}}\right)_{\alpha\alpha'} c^\pdagger_{\vec{k}\alpha'} \\ &\quad + \frac{1}{2} \sum_{\vec{k}} \left(c^\dagger_{\vec{k}\alpha} (\Delta_{\vec{k}})_{\alpha\alpha'} c_{-\vec{k}\alpha'}^\dagger + \text{H.c.}  \right). \label{MFHamiltonianCop}
\end{split}\end{align}
As in the main text, $h_{\vec{k}}$ and $\Delta_{\vec{k}}$ are the normal state Hamiltonian and superconducting order parameter matrix, respectively.

To probe the stability of the superconductor against impurity scattering, let us first consider a given disorder configuration,
\begin{equation}
\Delta H = \sum_{\vec{k},\vec{k}'} c^\dagger_{\vec{k}\alpha} \hat{W}_{\vec{k}\alpha,\vec{k}'\alpha'} c^\pdagger_{\vec{k}'\alpha'}, \label{GivenDisorderRealization}
\end{equation} 
with $\hat{W}^\dagger = \hat{W}$ due to Hermiticity. As is well-known, superconductivity behaves very differently in the presence of non-magnetic (time-reversal even, $t_W=+$) and magnetic (time-reversal odd, $t_W=-$) disorder.  Let us therefore split $\hat{W}$ into the respective components, $\hat{W} = \hat{W}^+ + \hat{W}^-$ with
\begin{equation}
 \hat{W}^{t_W}_{\vec{k}\alpha,\vec{k}'\alpha'} = t_W \, T_{\alpha\beta} \left(\hat{W}^{t_W}_{-\vec{k}\beta,-\vec{k}'\beta'}\right)^* T^\dagger_{\beta'\alpha'}. \label{DecompTRETRO}
\end{equation} 
Inspired by Anderson's work \cite{AndTh1}, we use a basis where Kramers partners, $c_{\vec{k}\alpha}$ and $T_{\alpha\beta}c^\dagger_{-\vec{k}\beta}$,  are manifest: defining the Nambu spinor $\Phi_{\vec{k}\alpha} = (c^\pdagger_{\vec{k}\alpha},T_{\alpha\beta}c^\dagger_{-\vec{k}\beta})^T$, the total Hamiltonian can be restated as $H_{0} +  \Delta H_{} = \frac{1}{2} \sum_{\vec{k},\vec{k}'} \Phi^\dagger_{\vec{k}\alpha} \left(\hat{h}^{\text{BdG}} \right)_{\vec{k}\alpha,\vec{k}'\alpha'}  \Phi_{\vec{k}'\alpha'}$. We split it into three parts, $\hat{h}^{\text{BdG}} = \hat{h}^{\text{BdG}}_n + \hat{h}^{\text{BdG}}_\Delta + \hat{h}^{\text{BdG}}_W$, defined by
\begin{align}\begin{split}
   \hat{h}^{\text{BdG}}_n &= \begin{pmatrix} \hat{h}  & 0 \\ 0 & -\hat{h}\end{pmatrix}, \\ \hat{h}^{\text{BdG}}_\Delta &= \begin{pmatrix} 0 & \hat{\mathcal{D}} \\ \hat{\mathcal{D}}^\dagger & 0 \end{pmatrix}, \\ \hat{h}^{\text{BdG}}_W &= \begin{pmatrix} \hat{W}^+ + \hat{W}^- & 0 \\ 0 & -\hat{W}^+ + \hat{W}^- \end{pmatrix},  \label{ExplBdGHams}
\end{split}\end{align} 
which correspond to the normal state Hamiltonian, the superconducting pairing, and the disorder potential, respectively. Here we use the same conventions, $\hat{h}_{\vec{k}\alpha,\vec{k}'\alpha'} = \delta_{\vec{k},\vec{k}'} \left(h_{\vec{k}}\right)_{\alpha\alpha'}$ and $(\hat{\mathcal{D}})_{\vec{k}\alpha,\vec{k}'\alpha'} = \delta_{\vec{k},\vec{k}'} (\Delta_{\vec{k}} T^\dagger)_{\alpha\alpha'}$, as in the main text. 

It is not difficult to see that the gap of the system is not reduced by the presence of disorder, $\hat{W}\neq 0$, if 
\begin{equation}
 \left[\hat{h}^{\text{BdG}}_n + \hat{h}^{\text{BdG}}_W,\,\hat{h}^{\text{BdG}}_\Delta \right]_+ = 0,  \label{ConditionForGapNotRed}
\end{equation} 
which indicates the stability of the superconductor against disorder. From \equref{ExplBdGHams} follows that  $[\hat{h}^{\text{BdG}}_n,\,\hat{h}^{\text{BdG}}_\Delta ]_+ = 0$ if the criterion in \equref{GATCond1} holds, i.e., if the normal state Hamiltonian and the superconducting order parameter commute. In the eigenbasis of the normal state Bloch Hamiltonian, this condition simply means
\begin{equation}
  (E_{\vec{k}l} - E_{\vec{k}l'}) \braket{\psi_{\vec{k}l}|\Delta_{\vec{k}}T^\dagger|\psi_{\vec{k}l'}} = 0, \label{SimplifiedCondition}
\end{equation}
where $h_{\vec{k}}\ket{\psi_{\vec{k}l}} = E_{\vec{k}l}\ket{\psi_{\vec{k}l}}$; in other words, all matrix elements of the order parameter between different bands $l$ and $l'$ with $\delta^{ll'}_{\vec{k}}=|E_{\vec{k}l}- E_{\vec{k}l'}| \neq 0$ have to be zero. This is a very natural assumption as it typically holds $|\Delta_{\vec{k}}| \ll \delta^{ll'}_{\vec{k}}$ in weak-coupling superconductors. However, in a generic basis, the condition \equref{GATCond1} has to be taken into account, as we discussed in the main text.

If $[\hat{h}^{\text{BdG}}_n,\,\hat{h}^{\text{BdG}}_\Delta ]_+ = 0$ holds, \equref{ConditionForGapNotRed} becomes $[\hat{h}^{\text{BdG}}_W,\,\hat{h}^{\text{BdG}}_\Delta]_+ = 0$, which we can be further simplified to
\begin{equation}
\sum_{t_W=\pm} \left[ \hat{W}^{t_W} ,  \hat{D} \right]_{-t_W} = 0.
\end{equation}
Focusing on non-magnetic and magnetic disorder separately, we have thus also recovered the second condition in \equref{GeneralizedATCond}. 

We finally note that the same result can be obtained within the more conventional diagrammatic approach \cite{AGD}: using the same low-energy description as in the main text where only states in the vicinity of the Fermi energy are included, it is shown in \refcite{Scheurer2016} that the superconducting critical temperature $T_c$ is not affect by disorder in leading order in $(k_F l)^{-1}$ if \equref{CommutatorInPSBasis} is satisfied. Diagrammatically, this results from a cancellation of the disorder-induced self energy and vertex correction. In the appendix below, we will recover this cancellation to leading order in the impurity strength. We emphasize again that such a low-energy description implicitly assumes that the first condition (\ref{GATCond1}) or, equivalently, \equref{SimplifiedCondition} is satisfied. 

\section{Limit of weak disorder}
\label{WeakDisorderLimit}
In this appendix, we will follow the more conventional approach and consider an ensemble of disorder configurations; all physical quantities, such as the free energy below, will be averaged over disorder realizations. We take the disorder configurations to be Gaussian distributed with zero mean such that their probability distribution is uniquely defined by the correlator
\begin{align}\begin{split}
    &\braket{\hat{W}_{\vec{x}_1\alpha_1,\vec{x}_1'\alpha_1'}\hat{W}_{\vec{x}_2\alpha_2,\vec{x}_2'\alpha_2'}}_\text{dis} \\& \quad = \delta(\vec{x}_1-\vec{x}_1')\delta(\vec{x}_2-\vec{x}_2')\delta(\vec{x}_1-\vec{x}_2) \Gamma_{\alpha_1 \alpha_1',\alpha_2\alpha_2'}. 
\end{split}\end{align}
Here $W_{\vec{x}_1\alpha_1,\vec{x}_1'\alpha_1'}$ is the real-space representation of the disorder potential $\hat{W}$ in \equref{GivenDisorderRealization}, $\braket{\dots}_\text{dis}$ denotes the average over disorder configurations, and $\Gamma_{\alpha_1 \alpha_1',\alpha_2\alpha_2'}$ encodes the orbital/spin structure of the impurities. In general, it can be expanded in Hermitian basis matrices $\{w_\mu\}$, 
\begin{equation}
    \Gamma_{\alpha_1 \alpha_1',\alpha_2\alpha_2'} = \sum_{\mu,\mu'} \gamma_{\mu \mu'} (w_{\mu})_{\alpha_1\alpha_1'} (w_{\mu'})_{\alpha_2\alpha_2'},
\end{equation}
allowing for the presence of different types of impurities at the same time \cite{OurDisorderSOC}. For notational simplicity, we will here focus on only one type of disorder at a time, $\Gamma_{\alpha_1 \alpha_1',\alpha_2\alpha_2'} = \gamma (W)_{\alpha_1\alpha_1'} (W)_{\alpha_2\alpha_2'}$. We will not specify the orbital structure of $W$ and only distinguish between time-reversal even ($t_W =+1$) and odd ($t_W=-1$) impurities, $\Theta W \Theta^\dagger = t_W W$. We normalize $W$ such that $\sum_{\alpha_1,\alpha_2} |W_{\alpha_1\alpha_2}|^2 = 2$.

The general expression for the disorder-averaged free energy $\braket{\mathcal{F}}_\text{dis}$ of \refcite{ScheurerPRB2016} is also valid in the pseudospin-triplet basis of the main text. Assuming, as usual, that the superconducting order parameter only depends on the position (label by $\Omega$ in the following) on the Fermi surface, we write $\Delta_{s}(\vec{k}) \equiv \Delta_{s}(\Omega)$ and find in leading order in the superconducting order parameter
\begin{align}
     \braket{\mathcal{F}}_\text{dis} \sim \sum_{s,s'}\int \mathrm{d} \Omega \int \mathrm{d} \Omega' \, \Delta^*_s(\Omega) D_{\Omega s,\Omega' s'}^{\phantom{-1}}(T) \Delta_{s'}(\Omega'). \label{AveragedFreeEnergy}
\end{align}
The kernel is given by
\begin{equation}
     D(T) = -T \sum_{\omega_n} \left( \mathcal{C}(\omega_n) - t_W \, \mathcal{S}^S \right)^{-1} -\mathcal{V}^{-1}, \label{ExpressionForD}
\end{equation}
where $\mathcal{S}_{\Omega s,\Omega' s'} = \mathcal{S}^0_{\Omega s,\Omega' s'} + \mathcal{S}^0_{\Omega s,\Omega'_{\text{K}} s'}$ (here $\Omega_{\text{K}}$ denotes the Kramers partner of $\Omega$) and $\mathcal{S}^0_{\Omega s,\Omega' s'} = |\braket{\phi_{\Omega}^s|W|\phi_{\Omega'}^{s'}}|^2$ with the chiral states $\ket{\phi_{\Omega}^s}$ defined in \equref{BasisChoice}. Furthermore, we have
\begin{align}\begin{split}
     &\mathcal{C}_{\Omega s,\Omega' s'}(i\omega_n) \\ &\quad= \frac{\delta_{s,s'}\delta_{\Omega,\Omega'}}{\rho_{\Omega s}} \left(\frac{|\omega_n|}{\pi} + \sum_{\tilde{s}} \int\textrm{d}\widetilde{\Omega} \, \rho_{\widetilde{\Omega}\tilde{s}} \, \mathcal{S}^S_{\Omega s,\widetilde{\Omega} \tilde{s}} \right),
\end{split}\end{align}
where $\rho_{\Omega s}$ is the angular-resolved density of states (within our current pseudospin approach with doubly-degenerate Fermi surfaces, it holds $\rho_{\Omega s} = \rho_{\Omega}$). Finally, the last term in \equref{ExpressionForD} is the inverse of the interaction kernel $\mathcal{V}_{\Omega s,\Omega' s'}$. While this term is crucial in determining the form of the superconducting order parameter, we will here \textit{assume} that a certain order parameter is realized (requiring a certain underlying $\mathcal{V}$) and study the impact of disorder on it. For that reason, $\mathcal{V}$ will not explicitly appear in the results below.

In \refcite{Scheurer2016}, it was shown that $\mathcal{S}$ cancels out entirely from \equref{ExpressionForD} if \equref{CommutatorInPSBasis} is satisfied. Here, we will focus on the leading order impact of disorder for a general superconductor for which the left-hand side of \equref{CommutatorInPSBasis} is non-zero. Straightforward asymptotic analysis of \equref{AveragedFreeEnergy} yields 
\begin{equation}
    \frac{T_c}{T_{c,0}} = 1 - \frac{\pi^2}{2 T_{c,0}}\rho_F \Gamma^{\text{eff}} +\mathcal{O}(\gamma^2),
\end{equation}
where $T_c$ ($T_{c,0}$) is the critical temperature in the presence (absence) of disorder, $\rho_F$ is the total density of states at the Fermi level, and the effective scattering rate 
\begin{align}\begin{split}
    \rho_F \Gamma^{\text{eff}} &= \frac{\gamma}{4\sum_s \int\mathrm{d}\Omega\, \rho_{\Omega s} |\Delta_s(\Omega)|^2}\sum_{s,s'}\int \mathrm{d}\Omega\mathrm{d}\Omega' \, \rho_{\Omega s}\rho_{\Omega' s'}  \\ &\quad \times\mathcal{S}^S_{\Omega s,\Omega' s'}\left(|\Delta_s(\Omega)|^2 - t_W \Delta^*_s(\Omega) \Delta_{s'}(\Omega') \right).
\end{split}\end{align}
Using the property \cite{SelectionRules} $\Delta_s(\Omega) = \Delta_s(\Omega_{\text{K}})$ that holds for any system with spinfull time-reversal symmetry, $\Theta^2= -\mathbbm{1}$, and re-expressing the integrals in terms of momentum averages ($N_{\Lambda}$ momentum points) over the states in the vicinity (cutoff $\Lambda$) of the Fermi energy,
\begin{equation}
    {\sum_{\vec{k}}}^{\text{FS}} \dots  := \frac{1}{N_\Lambda} \sum_n\sum_{\vec{k},|\epsilon_{\vec{k}n}|<\Lambda} \dots ,
\end{equation}
we can write
\begin{equation}
    \Gamma^{\text{eff}} \sim \gamma \frac{\sum_{\vec{k},\vec{k}'}^{\text{FS}}\sum_{s,s'} |C_{\vec{k}s,\vec{k}'s'}|^2}{4\sum_{\vec{k}}^{\text{FS}}\sum_s |\Delta_s(\vec{k})|^2}. \label{ExpressionForScatteringRate}
\end{equation}
Here $C_{\vec{k}s,\vec{k}'s'}$ is as defined in \equref{DefinitionOfCorrelatorInBasis} of the main text. Upon introducing the scattering rate $\tau^{-1} = 2\pi \rho_F \gamma$, we thus find \equref{SuppresionOfTcMainText} of the main text.

We point out that \equref{ExpressionForScatteringRate} can also be written in the manifestly basis-independent form
\begin{equation}
    \Gamma^{\text{eff}} \sim \gamma \frac{\sum_{\vec{k},\vec{k}'}^{\text{FS}}\text{tr}\left[ \hat{C}^\dagger_{\vec{k},\vec{k}'}\hat{C}^\pdagger_{\vec{k},\vec{k}'} \right] }{4\sum_{\vec{k}}^{\text{FS}}\text{tr}\left[\Delta_{\vec{k}}^\dagger \Delta_{\vec{k}}^\pdagger\right]},
\end{equation}
where the trace is over pseudospin space and
\begin{equation}
    \hat{C}_{\vec{k},\vec{k'}} =  \Delta_{\vec{k}} T^\dagger W_{\vec{k},\vec{k}'} - t_W W_{\vec{k},\vec{k}'}\Delta_{\vec{k}'}T^\dagger ;   \label{FinalPseudoSpinBasisCommutator}
\end{equation}
here all quantities are $2\times 2$ matrices in pseudospin space. 

Finally, note that we have chosen the normalization such that $\Gamma^{\text{eff}}=\gamma$ in the case of spin-magnetic impurities in a singlet superconductor with a constant gap in a one-band model (only spin, no orbital).
\begin{table*}[tb!]
    \centering
    \begin{tabular}{c|c|c|c|c|c|c|c|c}
    \hline
         Dose [C/cm$^2$] & $n_e$ [m$^{-3}$] & $n_h$ [m$^{-3}$] & $\mu_e$ [m$^2$/Vs]& $\mu_h$ [m$^2$/Vs] & $\tau_e^{-1}$ [meV] & $\tau_h^{-1}$ [meV] & $\ell_e$ [nm] & $\ell_h$ [nm]  \\
         \hline \hline
         0 & $4.2(1) \times 10^{27}$ & $2.2(1) \times 10^{27}$ & 0.10(1) & 0.28(1) & 70(2) & 26(2) & $344(5)$ & $725(5)$ \\
         1.33 & $4.2(1) \times 10^{27}$ & $2.2(1) \times 10^{27}$ & 0.05(1) & 0.14(1) &  140(2) & 53(2) & 172(5) & 363(5) \\
         \hline
    \end{tabular}
    \caption{Parameters obtained from fitting longitudinal and Hall resistivity at $T=5$~K to semiclassical two-band model expressions. For simplicity, the (inverse) scattering times are obtained under the assumption that $m^*=m_e$. For an effective mass different from the bare electron mass, they scale like $\tau_\alpha \rightarrow \tau_\alpha m_e/m^*$. Note that the mean-free paths $\ell_\alpha$ are independent of the effective masses.  }
    \label{tab:two-band_fit}
\end{table*}

\section{Two-band model fit to longitudinal and Hall resistivity}
\label{sec_app:two_band_model_fit_resistivity}

We fit the longitudinal and Hall resistivity $\rho_{xx}$ and $\rho_{xy}$ to the standard semiclassical expressions for a system with electron and hole charge carriers~\cite{Pippard}
\begin{align}
    \rho_{xx} &= \frac{1}{e_0} \frac{n_e \mu_e + n_h \mu_h + \mu_e \mu_h B^2 (n_e \mu_h + n_h \mu_e) }{(n_e \mu_e + n_h \mu_h)^2 + \mu_e \mu_h B^2 (n_e - n_h)^2} \\
    \rho_{xy} &= \frac{B}{e_0} \frac{n_h \mu_h^2 - n_e \mu_e^2 + (n_h - n_e) \mu_e^2 \mu_h^2 B^2}{(n_e \mu_e + n_h \mu_h)^2 + (n_e - n_h)^2 \mu_e^2 \mu_h^2 B^2}\,.
\end{align}

From the fit, we obtain the electon, $n_e$, and hole, $n_h$, charge carrier densities as well as their respective mobilities $\mu_e$ and $\mu_h$. This allows us to estimate the scattering rates $\tau_{e}^{-1}, \tau_{h}^{-1}$ and the mean-free paths $\ell_{e}, \ell_h$ in the system. All results from the two-band fit are collected in Table~\ref{tab:two-band_fit}.

\section{Dimensionless scattering rate and Abrikosov-Gorkov law}
\label{sec_app:dimensionless_scattering_rate}
Let us briefly describe our choice of using the dimensionless scattering rate 
\begin{equation}
    \gamma^\lambda = \frac{\hbar}{2 \pi k_B \mu_0 } \frac{\Delta \rho_0}{\lambda_0^2 
    T_{c,0}} = 0.97 \frac{\Delta \rho_0 [\mu \Omega \, \text{cm}]}{ \lambda_0^2 [10^{-7}\text{m}] T_{c,0} [\text{K}]}
\end{equation}
when comparing our experimental results of $T_c$ suppression with the Abrikosov-Gorkov (AG) law (see Fig~\ref{Tcgamma}). Here, $\Delta \rho_0 = \rho_0^{(\text{irradiated})} - \rho_0^{(\text{pristine})}$ denotes the change of the residual resistivity in the normal state induced by electron irradiation, $T_{c,0}$ is the transition temperature in the pristine sample, and $\lambda_0$ is the $T=0$ London penetration depth in the prisine sample. We find $T_{c,0} = 1.76$~K and $\lambda_0 = 220$~nm for PdTe$_2$.

The AG law relates the suppression of $T_c$ with the pair-breaking scattering rate $\tau_{pb}$ as
\begin{equation}
    \frac{\delta T_c}{T_{c,0}} = - \frac{\pi}{4 T_{c,0} \tau_{pb}} \,.
\end{equation}
For single-band, isotropic $s$-wave superconductors $\tau_{pb}$ is given by the magnetic (spin-flip) scattering rate $\tau_m$. Under the assumption that all scattering processes are pair-breaking $\tau = \tau_{pb}$, i.e.~purely magnetic disorder in the isotropic single-band $s$-wave case, the AG law suppression corresponds to $\zeta = 1$ in Eq.~\eqref{eq:Tc_suppression}. In general, not all scattering processes that contribute to the residual normal state resistivity $\rho_0 = m^*/(n e_0^2 \tau)$ are pair breaking, which leads to $\zeta < 1$. For example, in the case of non-magnetic (TRS) disorder, we have illustrated the pair-breaking processes for each of our candidate pairing states $A_{1g}^{++}, A_{1g}^{+-}, A_{1u}^{++}$ and $A_{1u}^{+-}$ in Fig.~\ref{fig:ScatteringDifferentSCs}. 

We use electron irradiation to tune the scattering time $\tau$ by creating point-like, non-magnetic defects in the material. We have explicitly shown that electron irradiation only affects the scattering time $\tau$ and does not change the carrier density (see Sec.~\ref{sec:results} and Table~\ref{tab:two-band_fit}). The change $\Delta \rho_0$ is thus directly proportional to $\Delta \tau^{-1}$, and thus
\begin{equation}
    \gamma^\lambda = \frac{\hbar}{2 \pi k_B \mu_0 } \frac{\mu_0 n e_0^2}{m^*} \frac{m^*}{n e_0^2} \frac{\Delta \tau^{-1}}{T_{c,0}} = \frac{\hbar \Delta \tau^{-1}}{2 \pi k_B T_{c,0}} \,,
\end{equation}
where we have used that the superfluid density equals the total carrier density at $T=0$. We can thus express the AG law using the dimensionless scattering rate $g_\lambda$, which can be experimentally measured, in the form
\begin{equation}
    \frac{\delta T_c}{T_{c,0}} = - \frac{\pi^2 g_\lambda }{2} \frac{\Delta \tau^{-1}_{pb}}{\Delta \tau^{-1}} \,.
\end{equation}
By plotting our experimental results of $\delta T_c/T_{c,0}$ versus $\gamma^\lambda$, we can extract the dimensionless parameter $\zeta = \frac{\Delta \tau^{-1}_{pb}}{\Delta \tau^{-1}}$ and compare with the AG law. The parameter $\zeta$ expresses the fraction of scattering processes that are pair breaking. For example, $\zeta = 1$ for purely magnetic (TRA) scattering in an isotropic, single-band spin-single superconductor. In contrast, $\zeta = 1/2$ for non-magnetic (TRS) disorder in the $A_{1g}^{+-}$ state of a two-band superconductor with $\rho_1 \Delta_1 = - \rho_2 \Delta_2$ (see Fig.~\ref{fig:A1gDetrimentalProcesses}), assuming that the ratio of interband to intraband scattering is one (corresponding to orbitally insensitive disorder). 

\section{Density-functional theory results}
\label{sec_app:DFT}
In this section, we discuss additional \emph{ab initio} results on the band structure, Fermi surface contour, and wavefunction characters in PdTe$_2$.
We also provide details on the density functional theory method we use. 

\subsection{Details of DFT method}
We perform calculations using our recently-developed \textit{ab initio} tight-binding (TB) framework~\cite{ke2019prb}.
Realistic TB Hamiltonians are constructed via the maximally localized Wannier functions (MLWFs) method~\cite{marzari1997prb} as implemented in \textsc{wannier90}~\cite{mostofi2014cpc} through a postprocessing procedure~\cite{marzari1997prb,souza2001prb,marzari2012rmp} using the output of the self-consistent Density functional theory (DFT) calculations.
The details of our methods and applications can be found in Ref.~[\onlinecite{ke2019prb}].

DFT calculations are performed using a full-potential linear augmented plane wave (FP-LAPW) method, as implemented in \textsc{wien2k}~\cite{WIEN2k}.
The primitive cell contains one formula unit, and experimental lattice parameters~\cite{kim1990} are adopted.
The generalized gradient approximation of Perdew, Burke, and Ernzerhof~\cite{perdew1996} are used for the correlation and exchange potentials.
To generate the self-consistent potential and charge, we employed $R_\text{MT}\cdot K_\text{max}=8.0$ with Muffin-tin (MT) radii $R_\text{MT}=$ 2.5 and 2.3\,{a.u.}, for Pd and Te, respectively.
The calculations are performed with 1078 $k$-points in the irreducible Brillouin zone (BZ). They are iterated until charge differences between consecutive iterations are smaller than 1.0$\times 10^{-4} e$ and the total energy difference is lower than $0.01$~mRy.
Note that spin-orbit coupling is included in the Hamiltonian.

We construct the TB Hamiltonian by using 54 MLWFs, which correspond to $s$-, $p$-, and $d$-type orbitals for each of the three atoms in the unit cell. A real-space Hamiltonian $H({\bf R})$ with dimensions 54$\times$54 is constructed to accurately represent the band structures in the energy window of interest. We focus on six pairs of doubly-degenerated bands around $E_\text{F}$ and examine how their wavefunction characters evolve along the $k$ paths in BZ. 
\subsection{Fermi surface}
\begin{figure}[tbh]
    \centering
    \includegraphics[width=\linewidth]{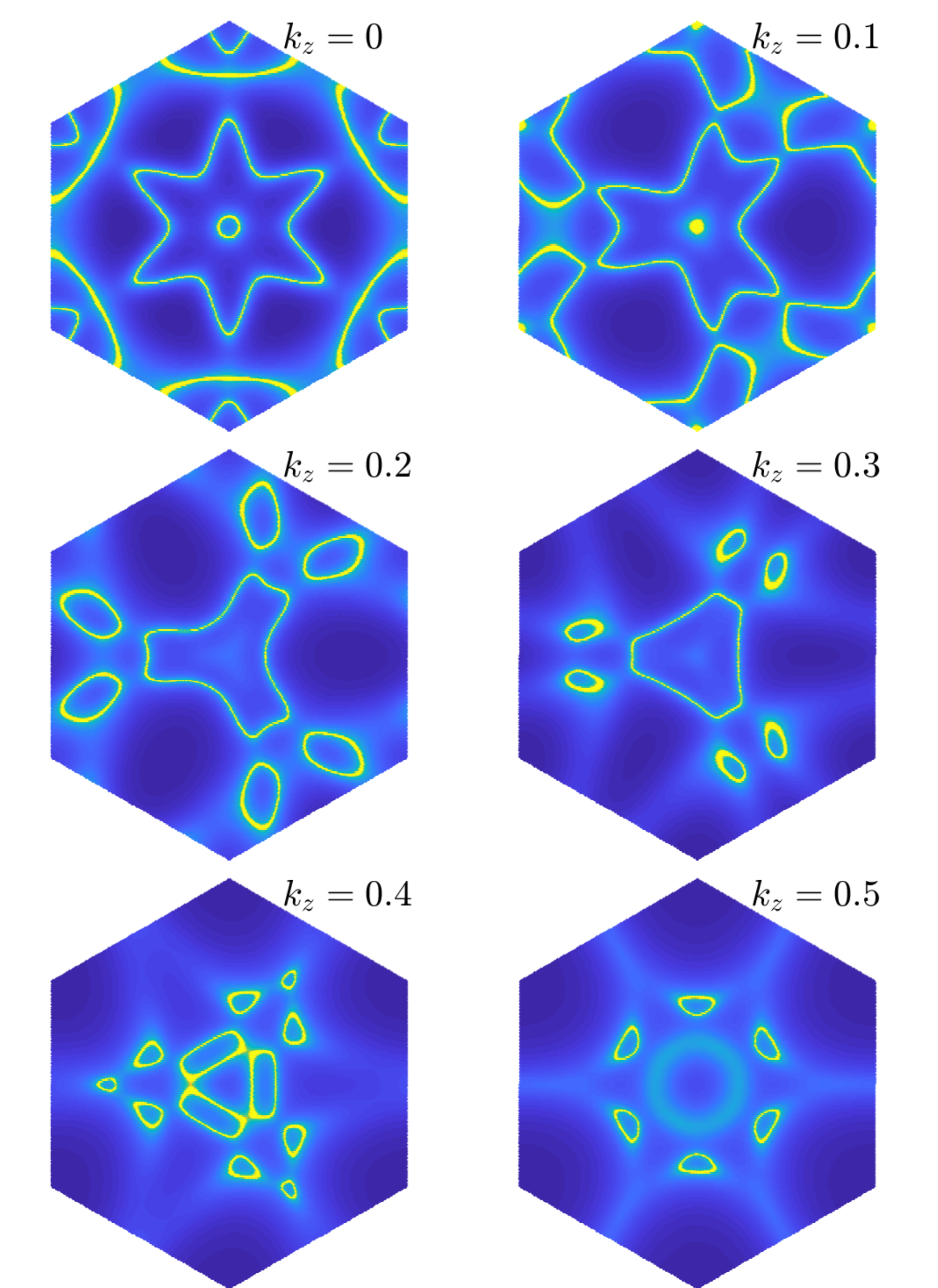}
    \caption{Fermi surface contour in PdTe$_2$ for
  $k_z=0;0.1;0.2;0.3;0.4;0.5$.}
    \label{fig:FS_contour_all}
\end{figure}

The band structure and a cross section of the Fermi surface in the  $k_{x}$-$k_{y}$ plane at $k_z = 0$ is shown in Fig.~\ref{fig:dft_main_text}. We provide additional Fermi surface cross sections for other values of $k_z$ in Fig.~\ref{fig:FS_contour_all}. The band structures are calculated in TB, within the energy window of interest, are essentially in perfect agreement with those obtained from DFT (not shown).
There are four pairs of doubly-degenerate bands across $E_\text{F}$ on the $\Gamma$--$M$--$K$ plane, forming two hole-pockets and two electron-pockets around $\Gamma$ and $K$, respectively.
The larger hole-pocket around $\Gamma$ has a strong anisotropy so that the $\Gamma$--$K$ direction has a much larger radius than the $\Gamma$--$M$ direction.

\subsection{Parity character of bands}
\begin{figure}[tbh]
    \centering
    \includegraphics[width=\linewidth]{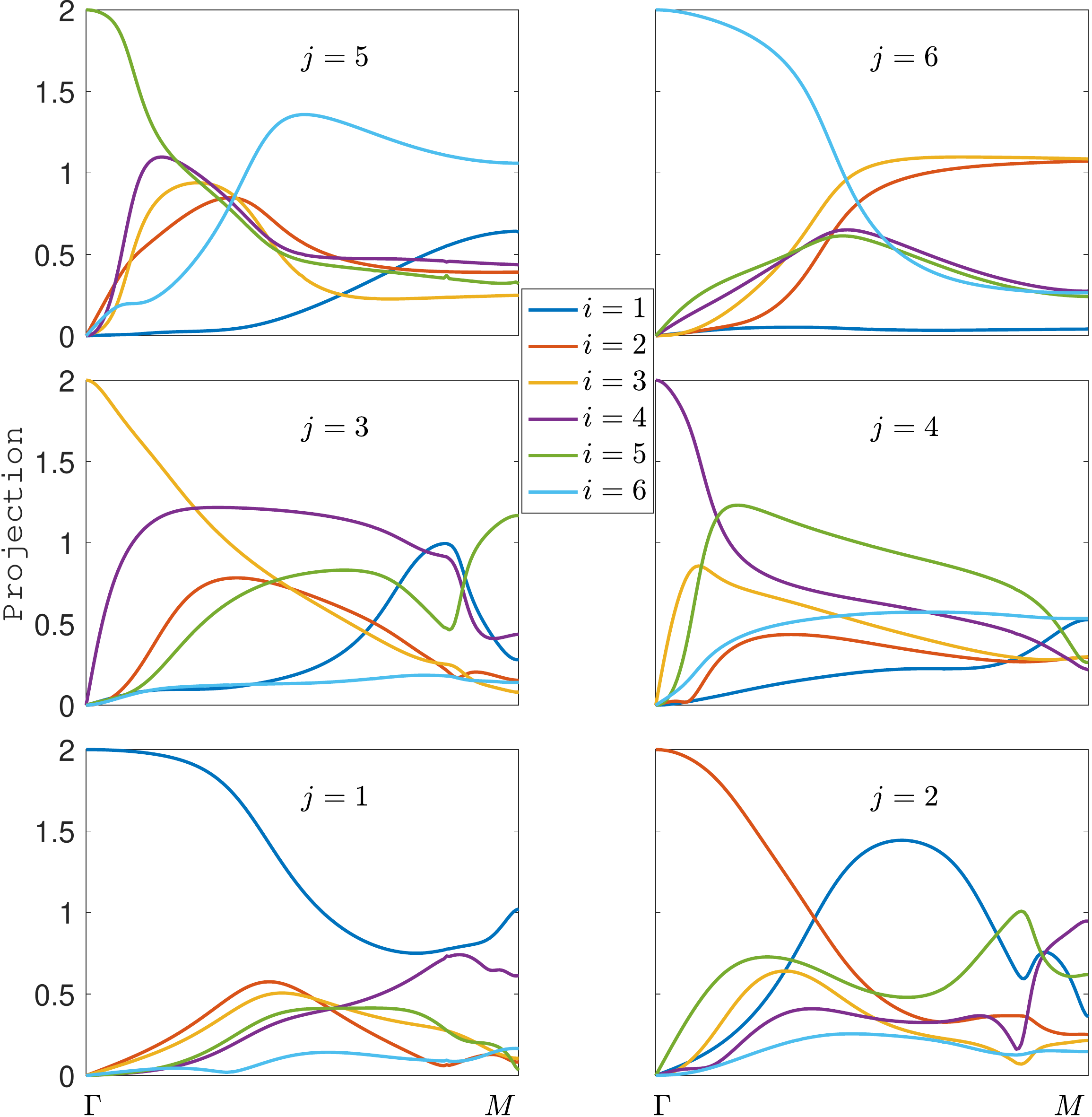}
    \caption{Projections of the eigenstates at momenta $\bf k$ along the $\Gamma$--$M$ line on those of the $\Gamma$ point, $P(\Gamma,i; {\bf k},j)$.
The index $i$ or $j$ denotes one of the six pairs of doubly-degenerated bands near $E_\text{F}$, as shown in Figs.~\ref{fig:dft_main_text} and~\ref{fig:band_structure_parities}. The index $i,j=1$ corresponds to the lowest energy band  and $i,j=6$ to the highest at $\Gamma$. }    \label{fig:dft_projections}
\end{figure}

\begin{figure}
    \centering
    \includegraphics[width=.8\linewidth]{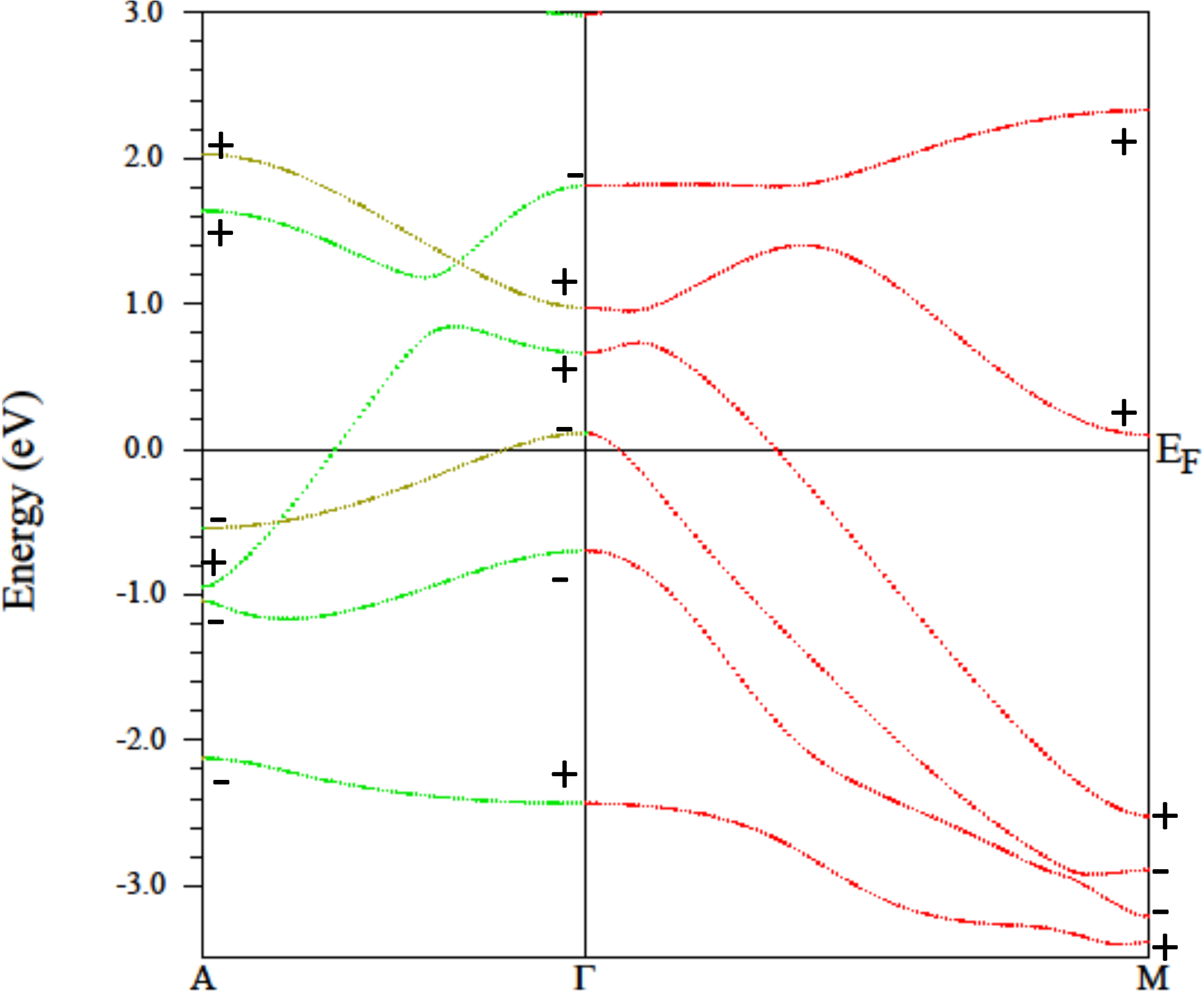}
    \caption{Band structure along A--$\Gamma$--M path in the Brillouin zone. The $\pm$ signs indicate the parity eigenvalues of the bands at the inversion symmetric high symmetry points. }
    \label{fig:band_structure_parities}
\end{figure}

We qualitatively estimate the parity mixing effects by calculating how the  characters of bands near $E_\text{F}$ change through the first BZ within TB.
The eigenvectors of ${\bf k}$ point are calculated and projected on those of the $\Gamma$ point, which have a well defined parity (see \figref{fig:band_structure_parities}). 
Figure~\ref{fig:dft_projections} shows the  calculated projections.
The projection of the $j_\text{th}$ pair at $\bf k$ on the $i_\text{th}$ pair at $\Gamma$ is defined as below
\begin{align}
P(\Gamma,i;{\bf k},j) &= \sqrt{\left|\langle\psi_{\Gamma}^{i,1}|\psi_{{\bf k}}^{j,1}\rangle\right|^2 + \left|\langle\psi_{\Gamma}^{i,2}|\psi_{{\bf k}}^{j,1}\rangle\right|^2}
  \\ &+ \sqrt{\left|\langle\psi_{\Gamma}^{i,1}|\psi_{{\bf k}}^{j,2}\rangle\right|^2 + \left|\langle\psi_{\Gamma}^{i,2}|\psi_{{\bf k}}^{j,2}\rangle\right|^2}.    
\end{align}
Here $i$ and $j$ denote the pair index of the six doubly-degenerated bands near $E_\text{F}$, and $\psi_{\bf k}^{i,1}$ and $\psi_{\bf k}^{i,2}$ are the wavefunctions of the two degenerated states of the corresponding $i_\text{th}$ pair at momentum $\bf k$.


\section{Gap anisotropies consistent with $T_c$ suppression}
\label{app_sec:gap_anisotropies}
In this section, we provide details of the behavior of the sensitivity parameter $\zeta$, which governs the $T_c$ suppression rate, for the case of multiple constant gaps on different Fermi sheets. We analyze Eq.~\eqref{SecondExpressionForZeta} in the case of $N=2$ (one gap ratio) and $N=4$ (three gap ratios). 

\subsection{Case of one gap ratio}
Here we consider the case of two different gap sizes $\Delta_1$ and $\Delta_2$ in the system. The gaps are assumed to be constant around a given Fermi sheet. This corresponds to the $N=2$ case in Eq.~\eqref{SecondExpressionForZeta}.
\begin{table}[t!]
    \centering
    \begin{tabular}{c|c|c|c}
    \hline
    	$C_1$ & $\nu$ & ($\Delta_1/\Delta_2)_1$ & ($\Delta_1/\Delta_2)_2$ \\
    	\hline \hline
    	$\{a\}$ &  $7.5\times 10^{-3}$ & $5.6$ & $-3.3$ \\
    	\hline
    	$\{b\}$ & $0.40$ & $2.1$ & $0.32$ \\
    	\hline
    	$\{c\}$ & $2.1$ & $2.8$ & $0.48$ \\
    	\hline
    	$\{d\}$ & $0.034$ & $3.3$ & $-0.98$ \\
    	\hline
    	$\{a,b\}$ & $0.41$ & $2.1$ & $0.33$ \\
    	\hline
    	$\{a,c\}$ & $2.1$ & $2.8$ & $0.48$ \\
    	\hline
    	$\{a,d\}$ & $0.042$ & $3.1$ & $-0.78$ \\
    	\hline
    	$\{c,d\}$ & $2.4$ & $3.0$ & $0.47$ \\
         \hline
    \end{tabular}
    \caption{List of all possible gap anisotropies $\Delta_1/\Delta_2$ distributed over the four Fermi sheets of PdTe$_2$ $\{a,b,c,d\}$ that are consistent with the experimentally observed sensitivity parameter $\zeta \simeq 1/16$. The gap ratios are obtained from Eq.~\eqref{SecondExpressionForZeta}. The set $C_1$ denotes the Fermi sheets with gap $\Delta_1$, the complementary set exhibits a gap of size $\Delta_2$. The value $\nu = \sum_{n \in C_1} \rho_n/(\sum_n \rho_n - \sum_{n \in C_1} \rho_n)$ denotes the ratio of the density of states on the two pockets with either $\Delta_1$ or $\Delta_2$, where we obtain $\rho_a = 0.01~\text{eV}^{-1}$, $\rho_b = 0.39~\text{eV}^{-1}$, $\rho_c = 0.91~\text{eV}^{-1}$, and $\rho_d = 0.05~\text{eV}^{-1}$ using DFT. We observe that the smallest anisotropy is obtained for the $A_{1g}^{+-}$ state with $C_1 = \{d\}$ corresponding to a sign change between the small electron pocket $d$ and the other three pockets. The minimal anisotropy for the $A_{1g}^{++}$ is about $2.1$ and is realized for various combinations $C_1 = \bigl( \{a,b\}, \{b\}, \{c\}, \{a,c\}, \{c,d\} \bigr)$. }
    \label{tab:gap_anisotropies}
\end{table}
 As shown in Fig.~\ref{fig:dft_main_text}, the Fermi surface manifold of PdTe$_2$ consists of four Fermi sheets with respective density of states 
 \begin{align}
     \rho_a& = 0.01 \; & \rho_b &= 0.39 \\
     \rho_c &= 0.91 \; & \rho_d &= 0.05 \,.
 \end{align}
 Let us denote the total Fermi surface manifold by $C = \{a, b, c, d \}$ and the subset that exhibits a gap $\Delta_1$ by $C_1$. The remaining set $C_2 = C \setminus C_1$ correspond to the sheets with a gap $\Delta_2$. In Table~\ref{tab:gap_anisotropies}, we present results of the gap anisotropies $\Delta_1/\Delta_2$ that are consistent with the experimentally observed sensitivity parameter $\zeta \simeq 1/16$. We consider all possible cases of how $\Delta_1$ and $\Delta_2$ are distributed over the four Fermi surfaces $a,b,c,d$. The table also contains the parameter $\nu$ that  enters Eq.~\eqref{SecondExpressionForZeta}. This is the ratio between the combined densities of states of the bands $C_1$ and $C_2$:
\begin{equation}
    \nu = \frac{\sum_{n \in C_1} \rho_n}{\sum_{n \in C_2} \rho_n} \,.
\end{equation}

 As shown in Table~\ref{tab:gap_anisotropies}, the most isotropic state occurs for the $A_{1g}^{+-}$ state with $C_1 = \{d\}$. For this state that two gap sizes are about equal in magnitude $|\Delta_1|/|\Delta_2| = 0.98$. This state is thus perfectly consistent with both the $T_c$ suppression rate we observe and the fact that the London penetration depth can be well captured by a single gap energy scale. 
 
  The smallest anisotropy we find for the $A_{1g}^{++}$ states is  about two: $\Delta_1/\Delta_2 = 2.1$ and $\Delta_1/\Delta_2 = 0.48$ (note that $1/0.48 = 2.1$). Such a state is realized for various ways of distributing $\Delta_1$ and $\Delta_2$ over the Fermi sheets
 \begin{align}
 \label{eq:C1_solutions}
     C_1 = \Bigl( \{a,b\}, \{b\}, \{c\}, 
\{a,c\}, \{c,d\} \Bigr) \,.
 \end{align}
 All other states that are consistent with a sensitivity parameter of $\zeta \simeq 1/16$ exhibit a degree of anisotropy larger than two. 

\subsection{General case of three gap ratios}
Here, we analyze the general case of $N=4$ in detail, where we allow for four different gap values on the four Fermi sheets $\{a,b,c,d\}$ of PdTe$_2$. As shown below, we find that the conclusions obtained from the $N=2$ case discussed above remain unchanged.  

For $N=4$, the expression for the sensitivity parameter in  Eq.~\eqref{SecondExpressionForZeta} takes the form
\begin{equation}
    \zeta = \frac12 - \frac{(1 + \sum_{j=a}^c \nu_j \eta_j)^2}{2 (1 + \sum_{j=a}^c \nu_j) (1 + \sum_{j=a}^c \nu_j \eta_j^2)}\,,
    \label{eq:zeta_N_4}
\end{equation}
where $\nu_j = \rho_j/\rho_d$ and $\eta_j = \Delta_j/\Delta_d$. Note that the summations run over the three Fermi surfaces $\{a, b, c \}$. We can use Eq.~\eqref{eq:zeta_N_4} to eliminate one of the $\eta_j$, say $\eta_c$, and obtain a family of solutions as a function of $\eta_a$ and $\eta_b$ that fulfill the condition $\zeta \simeq 1/16$ imposed by our experimental results. There are two independent solutions, $\eta^{\pm}_c(\eta_a, \eta_b)$, that differ from the sign in front of the square root ($\pm$). They need to be  investigated separately. 

To find the gap ratios that correspond to the most isotropic solutions, we minimize the function
\begin{equation}
\label{eq:height_function}
    h(\eta_a, \eta_b) = (\eta_a^2 - 1)^2 + (\eta_b^2 - 1)^2 + \bigl[\eta_{c}(\eta_a, \eta_b)^2 - 1\bigr]^2 \,.
\end{equation}
For $\eta^+_3$ there occur three local minima. The global minimum is very close to the $C_1 = \{d\}$ isotropic $A_{1g}^{+-}$ solution found above (see Table~\ref{tab:gap_anisotropies}):
\begin{equation}
    \eta_1 = 1.00\,,\; \eta_2 = -1.01\,,\; \eta_3 = -1.02\,.
\end{equation}
One of the other two local minima corresponds to a $(++++)$ solution, where the gap has the same sign on all the four Fermi surfaces: $\eta_1 = -1.00$, $\eta_2 = 0.33$, $\eta_3 = 0.96$. This state, however, has a larger degree of anisotropy ($\approx 3.0$) than the one found ($\approx 2.1$) for one gap ratio (see Eq.~\eqref{eq:C1_solutions}).

Turning to the analysis of the other solution $\eta_c^{-}$, we find three local minima of Eq.~\eqref{eq:height_function}. Two of them correspond to sign changing solutions with a larger degree of anisotropy than the state $C_1 = \{d\}$. The third local minimum, however, corresponds to a sign preserving $(++++)$ that is (slightly) more isotropic than any of the states found for one gap ratio. Specifically, its gap configuration reads
\begin{equation}
    \eta_a = 1.00\,,\; \eta_b = 1.07 \, , \; \eta_c = 0.51\,.
\end{equation}
Since $1/0.51 = 1.98 < 2.10$ this state is slightly less anisotropic than the solutions described by Eq.~\eqref{eq:C1_solutions}. Since the reduction of the degree of anisotropy is about $6\%$ only, however, our main conclusion that the sign preserving solutions are not consistent with the fact that $\lambda(T)$ can be described by a single gap energy scale still holds.

%%%%%%%%%%%%%%%%%%%%%%%%%%%% BIBLIOGRAPHY

%\vspace{2em}

%\newpage

%==================================================================================================================================

\end{document}